\documentclass[final,5p,times,twocolumn]{elsarticle}

\usepackage{amssymb}
\usepackage{amsmath}
\usepackage{bm}
\usepackage{amsthm}
\usepackage{graphicx}
\usepackage{extarrows}
\usepackage{graphicx}
\usepackage{xcolor}
\usepackage{ulem}
\usepackage[english=british]{csquotes} 

\journal{International Journal of Heat and Fluid Flow}

\begin{document}

\begin{frontmatter}



\title{Turbulent mesoscale convection in the Boussinesq limit and beyond}


\author[la1]{Shadab Alam}

\author[la1]{Dmitry Krasnov}

\author[la2]{Ambrish Pandey}

\author[la3]{John Panickacheril John}

\author[la1]{Roshan J. Samuel}

\author[la4]{Philipp P. Vieweg}

\author[la1,la5]{Jörg Schumacher}

\affiliation[la1]{organization={Institute of Thermodynamics and Fluid Mechanics},
             addressline={Technische Universität Ilmenau, P.O.Box 100565},
             city={Ilmenau},
             postcode={D-98684},
             country={Germany}}

\affiliation[la2]{organization={Department of Physics},
             addressline={Indian Institute of Technology Roorkee},
             city={Roorkee},
             postcode={247667},
             country={India}}

\affiliation[la3]{organization={Department of Aerospace Engineering},
             addressline={The University of Alabama},
             city={Tuscaloosa},
             postcode={35487},
             state={AL},
             country={USA}}

\affiliation[la4]{organization={Department of Applied Mathematics and Theoretical Physics},
             addressline={Cambridge University, Wilberforce Rd.},
             city={Cambridge},
             postcode={CB3 0WA},
             country={United Kingdom}}

\affiliation[la5]{organization={Tandon School of Engineering},
             addressline={New York University},
             city={New York City},
             postcode={11201},
             state={NY},
             country={USA}}

\begin{abstract}
Mesoscale convection covers an intermediate scale range between small-scale turbulence and the global organization of the convection flow. It is often characterized by an order of the convection patterns despite very high Rayleigh numbers and strong turbulent fluctuations. In this review, we discuss several aspects of mesoscale convection, which have been investigated by three-dimensional direct numerical simulations. The numerical studies are performed in a characteristic configuration of a plane layer that is heated from below and cooled from above or subject to constant heat flux at the top and bottom boundaries. We discuss the role of the thermal and mechanical boundary conditions for structure formation and study the impact of the domain shape as well as the Prandtl number. With respect to the latter, we focus on low values that arise in astrophysical convection and are partly not anymore accessible in laboratory experiments with liquid metals. Beside these experiments in the Boussinesq approximation, we report studies of non-Boussinesq mesoscale convection. This is done by investigating effects of compressibility and temperature dependence of material properties. The kinetic energy dissipation rate turns out to remain a central quantity for the turbulent mixing in compressible convection. Their different components, statistics, relation to the turbulent viscosity, and the multifractal properties are discussed.         
\end{abstract}

\end{frontmatter}

\section{Introduction}
\label{sec1}
The classical picture of fluid turbulence which exists since the seminal works by A. N. Kolmogorov, G. I. Taylor and L. Prandtl in the middle of the past century is that turbulence is characterized by a cascade of vortices, swirls and plumes of different size that give rise to a featureless and fully chaotic fluid motion. However, many natural flows in horizontally extended domains of size $L$ are organized in a hierarchy of regular prominent large-scale patterns, an ``order`` at an intermediate scale range. These coherent patterns are also denoted as long-living large-scale flow structures (LLFSs). They are observed even though the flows are highly turbulent with very high Reynolds or Rayleigh numbers, $Re$ and $Ra$, i.e., dimensionless measures of the vigour of the turbulence. In the case of turbulent thermal convection \cite{Kadanoff2001,Ahlers2009,Chilla2012}, two well-known generic examples for LLFSs exist.  

\begin{figure*}[ht]
\centering
\includegraphics[width=0.85\linewidth]{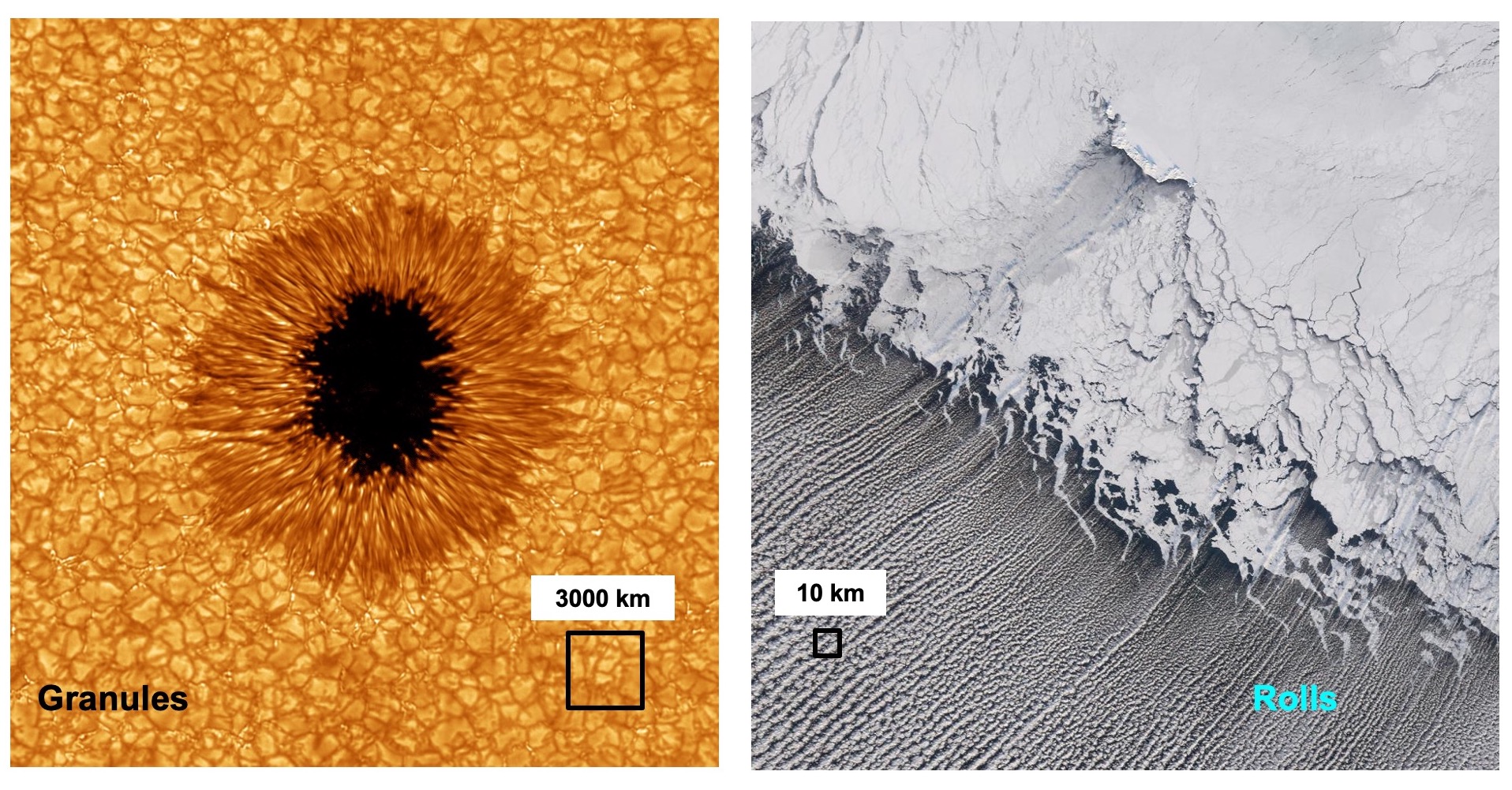}
\caption{Two prominent examples of mesoscale convection in nature. Left: solar granulation around a sunspot at the surface of Sun. Granules are the manifestation of the turbulent convection inside the star. Source: Swedish Solar Telescope (SST) at La Palma, Stockholm University. Observation taken 2002 by G. Scharmer. Right: cloud streets in the Bering Sea taken with the Moderate Resolution Imaging Spectroradiometer (MODIS) on the Terra satellite operated by NASA. Observation is taken on January 20th, 2006 (https://earthobservatory.nasa.gov/images/6243/cloud-streets-in-the-bering-sea). Parallel convection rolls form over the sea. In both figures, the grid spacing is indicated which is applied in global simulation models by a black box together with a typical mesh width in kilometers.}
\label{fig:Fig1}
\end{figure*}

The first one comprises \textit{granules and supergranules} in solar convection. These granules have a typical extension of about $10^3$ km and exist for about 10 minutes. They are the optically observable manifestation of convection that covers the solar surface as a regular network of cells, see Fig. \ref{fig:Fig1} (left). Warm fluid rises inside the cells to the top and sinks down at the speed of sound at its boundaries. Supergranules are about 30 times bigger and exist for about a day; they are not directly observable but can be traced indirectly by helioseismology or granule tracking \cite{Christensen2002}. One denotes the scales between $10^2$ km and $10^5$ km as the \textit{mesoscales} in this example. They should be compared to the depth of the convection zone with $H=2\times 10^5$ km (where temperature $T$ drops from 2 million K to 5779 K at the surface) and the solar circumference of $L\approx 4\times 10^6$ km, two global scales of solar convection. The Kolmogorov dissipation length $\eta_K$ can be estimated to about 1 cm; it is the smallest extension of vortices in solar convection \cite{Rincon2018,Schumacher2020}. Solar convection thus covers in total more than 11 orders of magnitude in scale. The dimensionless Rayleigh number is consequently very large with values of $Ra\sim 10^{18}$ to $10^{22}$. The second important parameter, the dimensionless Prandtl number $Pr$, which relates viscous to temperature diffusion in the ionized gas, is in contrast extremely small with $Pr\sim 10^{-6}$, caused by the dominance of photon-based energy transfer.

The second example comprises \textit{cloud streets and cellular patterns} in atmospheric convection. The LLFSs in atmospheric turbulence manifest as clouds which are arranged like pearls on a string up to heights of $H\approx 1$ km \cite{Markson1975}, see Fig. \ref{fig:Fig1} (right), as regular formations of open and filled cells \cite{Atkinson1996}, or as superclusters over oceanic warm pools for larger heights \cite{Mapes1993}. The mesoscale range starts here from about 1 km and extends up to 100 km or more. Cloud streets are reminiscent to the straight rolls of circulating fluid which are known from the onset of convection. These mesoscales should be compared again to a Kolmogorov dissipation length of 1 cm (or even smaller) and global scales of atmospheric motion starting from several thousand kilometers up to the circumference of the Earth with $L\approx 4\times 10^4$ km comprising again at least 8 orders of magnitude in scale. The Rayleigh number is again very large, $Ra\sim 10^{16}$ to $10^{18}$, the Prandtl number is $Pr\approx 0.7$. LLFSs cover again an intermediate range of lengths and times that extends itself now over approximately 2 to 3 orders of magnitude.  Figure \ref{fig:Fig1} displays both examples and indicates also the typical resolution of the corresponding global simulation grid, such as a climate model in the atmospheric example. A significant fraction of the meso-scale processes are of subgrid-scale character and thus have to be modeled and/or parametrized. 

Turbulent meso-scale convection (MC) displays a strong spatial coherence and order in the form of large-scale patterns that is absent at larger and smaller scales in the flow. These large-scale patterns in MC are considered as the essential link that drives global fluid motion at larger scales by an inverse and small-scale turbulence by a direct cascade. LLFSs affect thus the global transport of heat and momentum as well as turbulence statistics in multiple ways \cite{Pandey2018,Stevens2018,Green2020, Vieweg2021}. Their dynamical origin, typical scales and lifetimes, their dependence on boundary conditions, as well as their connection to extreme fluctuations at the microscales still needs to be better understood.  On the one hand, one expects that the large-scale patterns in MC will enhance the turbulent transport by providing the skeleton that carries momentum, heat, and mass across the convection layer. On the other hand, this coherence in the flow will feed a highly turbulent small-scale motion with statistical moments of derivatives of the velocity and temperature, that deviate strongly from Gaussian statistics \cite{Ishihara2009, Schumacher2018,Valori2021}.  The demand for more precise models of MC and their parametrization is necessary as motivated in Fig.  \ref{fig:Fig1}.

Mesoscale flows in nature are not only multiscale, but mostly also multi-physics phenomena, now termed \enquote{multi-X systems}. The convective turbulence inside the Sun interacts with generated magnetic fields and slow differential rotation. Close to the surface, the flow is compressible with strongly stratified profiles of the mean density, temperature and pressure, and with temperature-dependent material parameters. It is primarily driven by a strong outgoing radiative cooling flux \cite{Miesch2005,Nordlund2009}. Atmospheric MC couples to the transport of incoming solar radiation and the nonlinear thermodynamics of phase changes between vapor, liquid water and ice \cite{Stevens2005,Pauluis2011}. Both examples deviate strongly from the idealized Boussinesq limit which will be defined in subsection 2.5 \cite{Verma2018}. No existing numerical simulation model can incorporate all physical processes in their complex interplay and fully resolved. Nevertheless, non-Boussinesq extensions of MC -- all the way to the fully compressible regime of convection -- have to be considered for a complete understanding of the processes. Even resolving all scales of MC down to the dissipation lengths is impossible, but necessary to get deeper insights into the role of LLFSs as a driver of the highly non-Gaussian statistics at the smallest vortex sizes. Existing numerical models rely mostly on numerical or implicit viscosity and diffusion at the small scales \cite{Guerrero2016}. In turn, strongly fluctuating small-scale vortices and thermal plumes challenge all existing parametrizations of the unresolved scales, such as those by small-scale eddy viscosities and diffusivities in highly stratified flows \cite{Sreenivasan2019}. This is because high-amplitude, extreme derivatives of the fields are much more probable than for Gaussian statistics. 

\begin{figure}[h!] 
\centering
\includegraphics[width=0.99\linewidth]{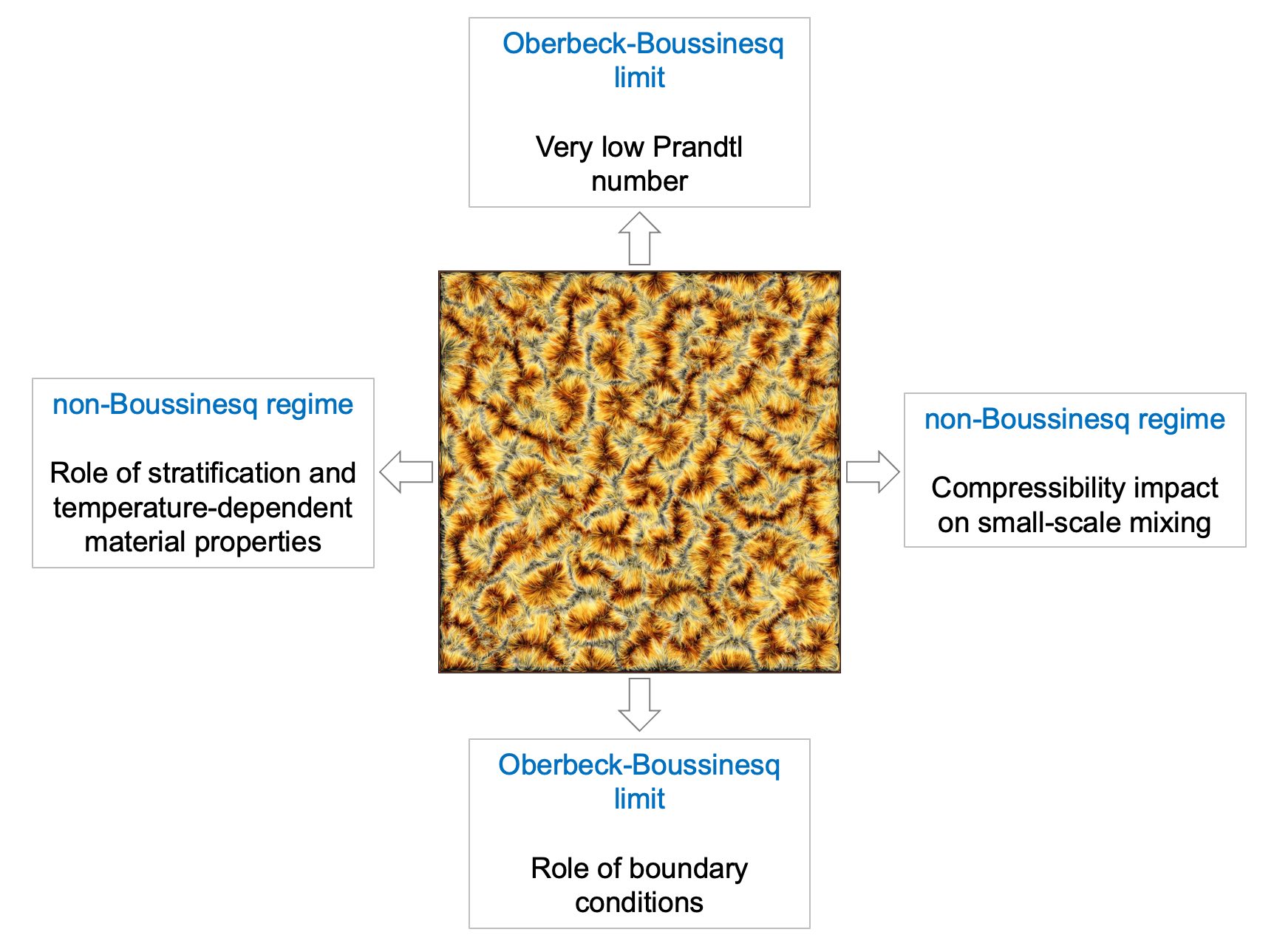}
\caption{Structure of the article in a diagram. Each subtopic, either in the Boussinesq or the non-Boussinesq regimes, is covered in a separate section. The role of very low Prandtl numbers in section 3, the role of thermal boundary conditions and domain shape for the structure formation in section 4, the role of stratification and temperature-dependent material properties in section 5, and the impact of compressibility on the small-scale mixing properties in section 6. }
\label{fig:Fig2}
\end{figure}

In this article, we summarize recent studies on mesoscale turbulent convection and present further results which extend these investigations. We are interested in several aspects of the connection between long-living large-scale flow structures, which include turbulent superstructures (TSSs) and supergranules (SGs) -- which will be differentiated in more detail in section \ref{sec4} --, and the statistical and turbulent transport properties of MC flows. As we have explained above, the complexity of MC in natural flows is too high, such that a deeper physical understanding requires to break down the complex flow into simpler building blocks of MC and understand the physical foundations there first. This strategy is illustrated in Fig. \ref{fig:Fig2}. We will thus report progress on several subtopics in the following sections, which detail one aspect each of the various facets of MC. 

The present investigations are built on three-dimensional direct numerical simulations (DNSs) of turbulent convection; they resolve the turbulence down to the viscous and diffusive scales and do not require any modeling of unresolved turbulence \cite{Moin1998}. We apply three simulation methods. For the Boussinesq cases, these are spectral element methods (SEMs), either with the CPU-based nek5000 \cite{Fischer1997,Scheel2013} or its GPU-accelerated successor nekRS \cite{Fischer2022}, which have been also used for the study of thermal convection processes in more complex geometries than plane layers \cite{Sachs2021,Vieweg2024b}. Furthermore, we applied a second-order finite difference method (FDM) for mesoscale convection studies at the smallest Prandtl numbers, which requires less memory in comparison to the spectral element method \cite{Krasnov2011}. The fully compressible flow cases are simulated with sixth-order compact finite difference schemes (CFDM) \cite{Baranwal2022}. These DNS methods are taken to understand the Navier-Stokes dynamics and its coupling to the temperature field in full detail. However they limit the range of accessible parameters in these studies. Clearly we will fall short by orders of magnitude when comparing Rayleigh, Reynolds and Prandtl numbers to the corresponding values in real flows in nature. 

The outline of the present manuscript is as follows. In section 2, we will outline the fully compressible equations of motion for the non-Boussinesq case and their simplification to the Boussinesq approximation limit. We list different boundary conditions and discuss the adiabatic and diffusive equilibria of the convection layer. Section 3 presents MC studies at very low Prandtl numbers together with a comparison at $Pr\sim 1$. Section 4 is dedicated to the role of boundary conditions on pattern formation and the effects of the domain shape on the characteristic pattern scales of LLFSs. Sections \ref{sec5} and \ref{sec6} are for compressible convection, a non-Boussinesq MC case. First, we give an overview of different regimes of compressible convection which depend on the degree of stratification of the adiabatic equilibrium  and superadiabaticity. The impact of an additional temperature dependence on material parameters follows then. Finally, we present results on the compressibility impact on turbulent small-scale mixing. This analysis is focused to the kinetic energy dissipation rate field in compressible convection. We conclude with a summary and an outlook in section \ref{sec7}.

\section{From fully compressible to Boussinesq convection}
\label{sec2}

\subsection{Equations of fully compressible convection} 

The equations of motion in the fully compressible flow, the most general case, comprise the balances of mass, momentum, and internal energy  densities which are denoted as $\rho$, $\rho u_i$, and $\rho e$, respectively. The equations are given by \cite{Landau1987}
\begin{align}
\frac{\partial \rho} {\partial t} &= - \frac{\partial \left(\rho u_{i}\right)}{\partial x_{i}} \,,
\label{eq:mass}\\
\frac{\partial\left(\rho u_{i}\right)}{\partial t} &= - \frac{\partial \left(\rho u_{i} u_{j} \right)}{\partial x_{j}}   -\frac{\partial p }{ \partial x_{i}} + \frac{\partial \sigma_{ij}}{\partial x_{j}} + \rho g_{i} \,,
\label{eq:mom}\\
\frac{\partial \left(\rho e\right)}{\partial t} &= - \frac{\partial \left(\rho e u_{j} \right)}{\partial x_{j}}   -p \frac{\partial u_{i} }{\partial x_{i}} +
\frac{\partial}{\partial x_{i}}\left(k \frac{\partial T}{\partial x_{i}}\right)
+ \sigma_{ij} S_{ij} \,,
\label{eq:ener}
\end{align}
with $i,j=1,2,3$. Variables $x_i$ stand for the spatial coordinates and $t$ for time. These balance equations are supplemented by an equation of state which connects pressure $p$, temperature $T$, and mass density $\rho$ and closes the system of equations. We assume that the fluid is an ideal gas. Then the equation of state  takes the following form
\begin{equation}
p(\rho, T) = R \rho T \,.
\label{eq:eos}
\end{equation}
Here, $e=c_v T$ with $c_v$ the specific heat at constant volume and $k$ the thermal conductivity in \eqref{eq:ener}. The equation of state \eqref{eq:eos} contains the specific gas constant $R=c_p-c_v$ with the specific heat at constant pressure is $c_p$ \index{Material parameters!specific heat at constant pressure}. The vector field of the acceleration due to gravity is given by $g_{i}=(0,0,-g)$ with $g=9.81$ m s$^{-2}$ on Earth. The viscous stress and the rate-of-strain tensor fields are defined as
\begin{align}
\sigma_{ij} &= \mu \left(\frac{\partial u_{i}}{\partial x_{j}} + \frac{\partial u_{j}}{\partial x_{i}} 
- \frac{2}{3} \delta_{ij} \frac{\partial u_{k}} {\partial x_{k}}\right)\,, 
\label{eq:vst}\\
S_{ij}  &=  \frac{1}{2}\left(\frac{\partial u_{i}}{\partial x_{j}} + \frac{\partial u_{j}}{\partial x_{i}} \right)\,.
\label{eq:st}
\end{align}
Both tensor fields are assumed to be symmetric. The dynamic or shear viscosity is $\mu$.  The material parameters are typically functions of the state variables, such that  $k(T,p)$ and $\mu(T,p)$. In most applications we have in mind, this simplifies to $k(T)$ and $\mu(T)$ together with the temperature dependence of the specific heat at constant pressure, $c_p(T)$. Frequently applied is the Sutherland law \cite{Sutherland1893} which gives $k\sim T^{3/2}$ and $\mu\sim T^{3/2}$ for the thermal conductivity and dynamic viscosity, respectively. Further details on the temperature dependence are discussed in section 5.

The velocity field can be decomposed into solenoidal ($s$) and dilatational ($d$) parts. This is known as the Helmholtz decomposition \index{Helmholtz decomposition} for any vector field:
\begin{equation}
{\bm u}={\bm u}_s({\bm x},t)+{\bm u}_d({\bm x},t) 
\label{Helmholtz}
\end{equation}
This decomposition implies that ${\bm \nabla} \cdot {\bm u}_{s} = 0$ and ${\bm \nabla} \times {\bm u}_{d} = 0$, and becomes important when the kinetic energy dissipation fields are introduced in section 6. A central reference velocity is the speed of sound which divides flows into sub- and supersonic turbulent flows; it is given by 
\begin{equation}
c_s=\sqrt{\frac{\partial p}{\partial \rho}\Big|_s}\,.
\label{sound}
\end{equation}
The derivative is taken at a constant specific entropy $s$.

\subsection{Boundary conditions in convection simulations}
\label{boundary_conditions}

The equations of motion for a given fluid flow problem require boundary and initial conditions. Periodic boundary conditions are the simplest ones to use. For any field or field component, $f=\{u_i, p, T, \rho\}$, one sets for example $f(x_1, x_2, x_3, t)=f(x_1+L, x_2, x_3, t)$ with the periodicity length $L$ (similar for the other space directions). Turbulent flows are, however, often enclosed by walls (w) and interfaces. For both the velocity and temperature field, there are two generic conditions that are frequently applied at such boundaries. 

No-slip boundary conditions for velocity field state that for all times $t$
\begin{equation}
u_i\big|_{\rm w}=0 \quad\text{for}\quad i=1,2,3 \,.  
\label{eq:noslip}
\end{equation}
Frequently used in the astro- and geophysical context are free-slip or stress-free boundary conditions for the velocity field. These boundary conditions state that the tangential viscous stresses vanish at the wall, which implies that
\begin{equation}
u_n\big|_{\rm w}=0 \quad\text{and}\quad \frac{\partial u_t}{\partial x_n}\Bigg|_{\rm w}=0\,.   
\label{eq:freeslip}
\end{equation}
Here, the index $n$ stands for the normal to the wall and $t$ for the tangent to the wall. This boundary condition is a rough approximation of a free non-deformable surface, which implies that the fluid can slip freely along the boundary. In the compressible flow case, particularly when the typical velocities exceed the speed of sound $c_s$, characteristic velocity boundary conditions have to be applied such that sound waves are not reflected at the boundary. This will not be the case for a purely buoyancy-driven compressible flow.

The Dirichlet boundary condition for the temperature field $T$ states that
\begin{equation}
T\big|_{\rm w}=T_0 \,.  
\label{Dirichlet} 
\end{equation}
Rather than fixing temperatures at the boundary, one may want to fix the (local) heat flux across the boundary. This results in the Neumann boundary condition for the temperature field which states that
\begin{equation}
\frac{\partial T}{\partial x_n}\Bigg|_{\rm w}=\beta_0 \,.  
\label{Neumann}
\end{equation}
A comparison of these boundary conditions with respect to the transfer of heat in convection has been conducted for example in refs. \cite{Verzicco2008,Johnston2009}. If $\beta_0=0$, the wall is thermally insulated or adiabatic. In general, $T_0$ and $\beta_0$ can be functions of the tangential coordinates at the boundary.  These thermal boundary conditions are idealizations that are typically hard to obtain in controlled laboratory experiments of MC, where the working fluid (fl) is controlled via plates (pl), which are themselves thermal conductors \cite{Moller2022,Weiss2023}. While no-slip conditions for the velocity field are well satisfied at the interface, the thermal conditions are generally given by matching conditions of temperature and (conductive) heat flux \cite{Vieweg2024b}, i.e., 
\begin{equation}
T_{\rm fl}\big|_{\rm w}=T_{\rm pl}\big|_{\rm w} \quad\text{and}\quad k_{\rm fl}\frac{\partial T_{\rm fl}}{\partial n}\Bigg|_{\rm w}=k_{\rm pl}\frac{\partial T_{\rm pl}}{\partial n}\Bigg|_{\rm w} \,,   
\label{conj}
\end{equation} 
where the thermal conductivities of the fluid and plate are given by $k_{\rm fl}$ and $k_{\rm pl}$, respectively. Equations \eqref{conj} are also referred to as the conjugate conditions \index{Boundary conditions!conjugate boundary conditions} on the plate-fluid interface \cite{Perelman1961}. The Dirichlet and Neumann boundary conditions from above can be considered as limits of the ratio of the thermal diffusivities, $\kappa_{\rm pl}/\kappa_{\rm fl}$: the Dirichlet case follows for $\kappa_{\rm pl}/\kappa_{\rm fl}\to \infty$, whereas the Neumann case follows for $\kappa_{\rm pl}/\kappa_{\rm fl}\to 0$. 
Alternatively, if the domain boundary is in contact with yet another fluid flow, this can be modelled using Newton (cooling) boundary conditions
\begin{equation}
\label{eq:BC_Newton_cooling}
k_{\textrm{fl or pl}} \left. \frac{\partial T}{\partial n} \right|_{\textrm{w}} = h_{\textrm{N}} \left( \left. T \right|_{\textrm{w}} - T_{\infty} \right)
\end{equation}
which reduce the potentially complex adjacent fluid flow to its undisturbed temperature $T_{\infty}$ and the manifesting heat transfer coefficient $h_{\textrm{N}}$. Note the similarities and differences between equations \eqref{conj} and \eqref{eq:BC_Newton_cooling}. Such Newton boundary conditions are helpful when mimicking laboratory experiments via digital twins \cite{Vieweg2024b} or in general when a non-trivial flow needs to be replaced by a simpler boundary condition.

\subsection{Adiabatic and diffusive equilibrium}

Equilibrium states are the reference configurations for convection with respect to which stability properties are evaluated; reference states are important for evaluating turbulent transport of mass, heat and momentum as well. In the following it can be seen that actually two equilibria can be obtained from the compressible equations of motion \eqref{eq:mass}--\eqref{eq:eos} \cite{Jones2022}. We discuss the case of Dirichlet boundary conditions in detail now. The \textit{ adiabatic equilibrium} state is obtained from the hydrostatic equilibrium condition \index{Equilibrium!hydrostatic} in which the fluid is at rest, $u_i=0$. From \eqref{eq:mom} one gets
\begin{equation}
\frac{d\bar{p}}{dx_3}=-g\bar{\rho}\,.
\label{eq:hydr_eq}
\end{equation}
We assume again an extended, planar, horizontal convection layer of height $H$ with a bottom boundary at $x_3=0$ and a top boundary at $x_3=H$. This implies that the equilibrium profiles of the thermodynamic state variables will be functions of the vertical coordinate $x_3$ only. These profiles are denoted by an overbar in the following. Together with the equation of state \eqref{eq:eos}, the assumption of a polytropic gas law, $p={\rm const}\times \rho^{\gamma}$ with $\gamma=c_p/c_v$, and a temperature difference of $\Delta\bar{T}=T_{\rm bot}-T_{\rm top}>0$, one gets
\begin{subequations}
\begin{equation}
\bar{T}(x_3)=T_{\rm bot} \left(1-\theta\frac{x_3}{H}\right)\,,
\label{eq:T_ad}
\end{equation}
\begin{equation}
\bar{\rho}(x_3)=\rho_{\rm bot} \left(1-\theta\frac{x_3}{H}\right)^{\frac{1}{\gamma-1}}\,,
\label{eq:rho_ad}
\end{equation}
\begin{equation}
\bar{p}(x_3)=\frac{g(\gamma-1) H\rho_{\rm bot}}{\gamma \theta} \left(1-\theta\frac{x_3}{H}\right)^{\frac{\gamma}{\gamma-1}}\,.
\label{eq:p_ad}
\end{equation}
\end{subequations}
We used $\theta=\Delta\bar{T}/T_{\rm bot}$ and that $\Delta\bar{T}=gH/c_p$. Note that the bottom values are the reference values. The dry adiabatic lapse rate is given by 
\begin{equation}
\frac{d\bar{T}}{d x_3}=-\frac{g}{c_p}\,.
\end{equation}
The adiabatic equilibrium state \index{Equilibrium!adiabatic} satisfies $d^2\bar T/d x_3^2=0$ and a constant entropy density across the layer, $ds/dx_3=0$. The exponent in \eqref{eq:rho_ad} is sometimes also rewritten as the polytropic index $m=(\gamma-1)^{-1}$.

The \textit{diffusive equilibrium} or pure heat conduction state is not necessarily isentropic. It satisfies boundary conditions for the entropy density, $s=s_{\rm bot}$ at $x_3=0$ and $s=s_{\rm top}$ at $x_3=H$. To avoid confusion, the diffusive equilibrium will be denoted by a tilde instead of a bar for the following.  The equilibrium conditions are as follows
\begin{equation}
\frac{d^2\tilde{T}}{d x_3^2}=0 \quad \mbox{and}\quad \frac{d\tilde{p}}{dx_3}=-g \frac{\tilde p}{R\tilde T}\,.
\label{eq:cond_di}
\end{equation}
One gets the following relations for the state variables,
\begin{subequations}
\begin{equation}
\tilde{T}(x_3)=T_{\rm bot} \left(1-\tilde{\theta}\frac{x_3}{H}\right)\,,
\label{eq:T_di}
\end{equation}
\begin{equation}
\tilde{\rho}(x_3)= \rho_{\rm bot} \left(1 - \tilde{\theta}\frac{x_3}{H}\right)^{\tilde m} \,,
\label{eq:rho_di}
\end{equation}
\begin{equation}
\tilde{p}(x_3)= \frac{gH \rho_{\rm bot}}{(\tilde m+1)\tilde\theta} \left(1 - \tilde{\theta}\frac{x_3}{H}\right)^{\tilde m+1}\,,
\label{eq:p_di}
\end{equation}
The boundary conditions $\tilde{T}(x_3=0)=T_{\rm bot}=\tilde{T}_{\rm bot}$ and  $\tilde{T}(x_3=H)=\tilde T_{\rm top}$ results again to a linear temperature profile of \eqref{eq:cond_di}.
Note that the bottom temperature is the same for both, the adiabatic and the diffusive equilibrium, i.e., $T_{\rm bot}=\bar T_{\rm bot}=\tilde{T}_{\rm bot}$. The exponent $\tilde{m}$ is given by
\begin{equation}
\tilde{m}+1= \frac{gH}{R T_{\rm bot}\tilde\theta} 
\end{equation}
\end{subequations}
The adiabatic temperature drop across the convection layer is smaller than the one for the diffusive equilibrium. 

\subsection{Superadiabaticity and strength of stratification}

The superadiabaticity $\varepsilon$ is a measure of the departure of the convective regime from the adiabatic equilibrium state and is defined as \cite{Verhoeven2015,Jones2022}
\begin{equation}
\varepsilon= - \frac{H}{T_{\rm bot}} \left[ \frac{d\tilde T}{\partial x_3}+ \frac{g}{c_p}\right] = \frac{H}{T_{\rm bot}} \Bigg| \frac{d\tilde T}{\partial x_3}\Bigg| - D\,,
\label{adia1}
\end{equation}
with the dissipation number $D$, a second central parameter next to $\varepsilon$ given by
\begin{equation}
D=\frac{gH}{c_p T_{\rm bot}}\,.
\label{Dnumber}
\end{equation}
Equation \eqref{adia1} can be rewritten as
\begin{equation}
\varepsilon= - \frac{H}{T_{\rm bot}} \left[\frac{\Delta \bar T}{H}-\frac{\Delta\tilde T}{H}\right] = \frac{\bar{T}_{\rm top}-\tilde{T}_{\rm top}}{T_{\rm bot}} >0\,.
\label{adia2}
\end{equation}
Equation \eqref{adia2} states that for small superadiabaticity, i.e., $\varepsilon\to 0$, diffusive and adiabatic equilibrium state coincide.

Which state is relevant as the reference state, the diffusive or the adiabatic equilibrium? For small Rayleigh numbers above the onset $Ra \gtrsim Ra_c$, the diffusive equilibrium is relevant. Then the definition
\begin{equation}
Ra=\alpha \Delta T\frac{gH^3}{\nu\kappa} = \frac{\Delta T}{T_{\rm bot}} \frac{gH^3}{\nu\kappa}\quad \mbox{with}\quad \Delta T= T_{\rm bot}-T_{\rm top} \,,
\label{ra1}
\end{equation}
can be used. We used that for an ideal gas $\alpha = 1/T_{\rm bot}$. For large Rayleigh numbers $Ra \gg Ra_c$ the adiabatic equilibrium has to be taken and the superadiabatic Rayleigh number $Ra_{\rm sup}$ should quantify the excess of the temperature drop over the one of the adiabatic state,
\begin{equation}
Ra_{\rm sup}=\varepsilon \frac{gH^3}{\nu\kappa} = \frac{\bar T_{\rm top} - T_{\rm top}}{T_{\rm bot}} \frac{gH^3}{\nu\kappa} \,.
\label{ra2}
\end{equation}
The superadiabaticity, which will be used here, is thus given by
\begin{equation}
\varepsilon  = \frac{\bar T_{\rm top} - T_{\rm top}}{T_{\rm bot}} \,.
\label{ra3}
\end{equation}
When the dry adiabatic lapse rate is small and we are well below scale height, we can substitute the numerator of \eqref{ra3} by $\Delta T\approx (T_{\rm bot} - T_{\rm top})$ in \eqref{ra3} which is consistent \eqref{ra1}.  For the following, we will drop the suffix ``sup'' in the definition of dimensionless parameters such as the Rayleigh number. Note that the superadiabaticity is not necessarily very small as in the anelastic case, as will be discussed in the next section.

\begin{figure}[t]
\centering
\includegraphics[width=0.8\linewidth]{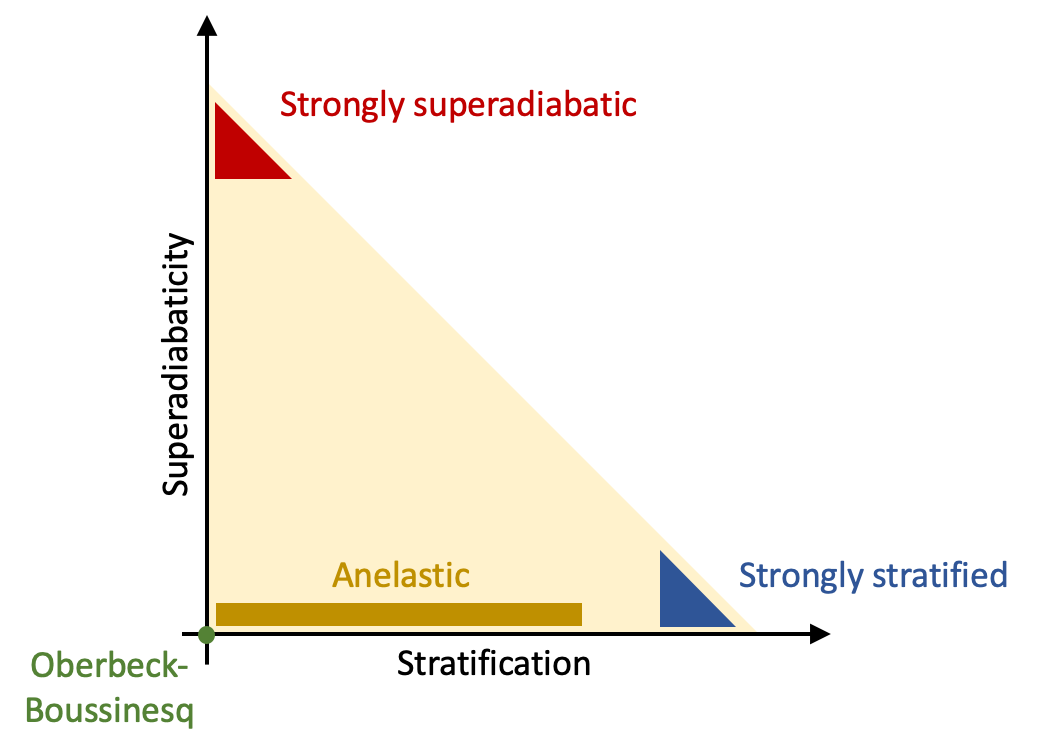}
\caption{Parameter plane of compressible convection spanned by superadiabaticity $\epsilon$ and dissipation number $D$, a measure of the strength of the stratification of the adiabatic temperature profile.} 
\label{fig:parameter_triangle}
\end{figure}

By definition both parameters, $D$ and $\epsilon$ can vary between 0 and 1. Furthermore, $\bar T_{\rm top}=\bar T(x_3=H)=T_{\rm bot}(1-D)$. Thus eqns. \eqref{Dnumber} and \eqref{adia2} can be combined to $\varepsilon +D = \Delta T/T_{\rm b}$ which leads to an inequality that relates both parameters
\begin{equation}
D \le 1-\varepsilon\,.
\label{adia3a}
\end{equation}
As a consequence both parameters span a triangular parameter space in which different regimes of compressible convection can be identified. This is illustrated in Fig. \ref{fig:parameter_triangle}. Indicated are the Boussinesq approximation at the origin of the parameter plane which is obtained for the limit $\varepsilon\to 0$ and the subsequent limit $D\to 0$. For the regime $\varepsilon \ll 1$ and moderate $D$, one is in the anelastic approximation. In this regime sound waves are filtered out. The continuation to $D\to 1-\varepsilon$ at $\varepsilon \ll 1$ results to the strongly stratified limit regime of convection while for $D\ll 1$ and $\varepsilon\to 1-D$ the strongly superadiabatic limit is obtained \cite{Panickacheril2023}, see section 5 for more details. In the following, the Boussinesq regime of turbulent convection will be detailed.

\subsection{Boussinesq approximation} 
 
In many convective flows the characteristic velocities are significantly smaller than the speed of sound, $u\ll c_s$; such flows remain close to their adiabatic equilibria. For approximately constant material properties $c_v$, $c_p$, $k$, and $\mu$, we can rewrite the left hand side of the energy balance equation \eqref{eq:ener} as an equation for the temperature field $T$,   
\begin{equation}
\frac{\partial \left(\rho e\right)}{\partial t} + \frac{\partial \left(\rho e u_{j} \right)}{\partial x_{j}}+ p \frac{\partial u_j}{\partial x_j} = 
 c_v\rho \left(\frac{\partial T}{\partial t}+ u_j\frac{\partial T}{\partial x_j}\right)-\frac{p}{\rho} \frac{{\rm D}\rho}{{\rm D}t}\,.
\end{equation}
Thus, one gets 
\begin{equation}
\rho c_p\frac{{\rm D}T}{{\rm D} t} - \frac{{\rm D}p}{{\rm D}t} = k \left(\frac{\partial^2 T}{\partial x_{i}^2}\right) + \sigma_{ij} S_{ij} \,,
\label{eq:ener1}
\end{equation}
as the new energy balance equation where $D/Dt$ stands for the material derivative. A simplification of the system of fully compressible eqns. \eqref{eq:mass}, \eqref{eq:mom}, and \eqref{eq:ener} together with the equation of state \eqref{eq:eos} follows for the limit $\varepsilon\to 0$ and a subsequent limit of  $D\to 0$. This regime is known as the Oberbeck-Boussinesq (OB) approximation \cite{Oberbeck1879,Boussinesq1903} which we will term sometimes as Boussinesq approximation for simplicity. Rewriting definition \eqref{Dnumber} as  
\begin{equation}
D=\frac{g}{c_{p}T_{\rm bot}} H = - \frac{H}{T_{\rm bot}} \frac{dT}{dr}\Bigg|_s = \frac{H}{{\cal H}_T}\,
\end{equation}
reveals that this limit corresponds to a layer height $H$ much smaller than the temperature scale height ${\cal H}_T$ which is given by 
\begin{equation}
{\cal H}_T= - \frac{dr}{d \log T}\,.
\label{scaleheight}
\end{equation}
The resulting Boussinesq equations of thermal convection in dimensionless form are given by  
\begin{align}
\frac{\partial u_i}{\partial x_i}&=0\,,
\label{eq:mass_bou}\\
\frac{\partial u_i}{\partial t}+ u_j \frac{\partial u_i}{\partial x_j} &= -\frac{\partial p^{\prime}}{\partial x_i} + 2\sqrt{\frac{Pr}{Ra}}\frac{\partial S_{ij}}{\partial x_j}  + T^{\prime} \delta_{i3}\,,
\label{eq:mom_bou}\\
\frac{\partial T^{\prime}}{\partial t}+ u_j \frac{\partial T^{\prime}}{\partial x_j} &= \frac{1}{\sqrt{Ra Pr}}\frac{\partial^2 T^{\prime}}{\partial x_j^2}\,.
\label{eq:ene_bou}
\end{align}
Here, the Rayleigh number $Ra$ is given by \eqref{ra1}. The Prandtl number $Pr$ is given by 
\begin{equation}
Pr=\frac{\nu}{\kappa}\,,
\end{equation}
with the kinematic viscosity $\nu$ and the thermal diffusivity $\kappa=k/(c_p \rho)$.
The velocity field is then incompressible and the temperature fluctuations enter directly the buoyancy forcing term on the right hand side of the momentum balance \eqref{eq:mom_bou}, a consequence of the simplified equation of state which reads now
\begin{equation}
\frac{\rho^{\prime}}{\bar{\rho}} =- \frac{T^{\prime}}{\bar{T}}\,. 
\label{eq:eos_bou}
\end{equation} 
The system of these equations is also known as the Rayleigh-B\'{e}nard convection (RBC) model. 

In the RBC setup, the Nusselt number $Nu$, a dimensionless number which characterizes the turbulent heat transfer, is obtained from \eqref{eq:ene_bou}. Let us assume no-slip boundary conditions at the top and bottom of a plane layer of height $H$ as well as periodic boundary conditions at the side faces. Using incompressibility, statistical stationarity, and averaging with respect to horizontal planes $A$ (at constant $x_3$) and time $t$ \citep{Otero2002}, one obtains
\begin{equation}
\frac{\partial \langle u_3T\rangle_{A,t}}{\partial x_3} -\frac{1}{\sqrt{Ra Pr}}\frac{\partial^2 \langle T\rangle_{A,t}}{\partial x_3^2}=0 \,.
\end{equation}
An additional integration with respect to $x_3$ shows that for each $x_3\in [0,H]$ the sum of the convective and diffusive heat current is constant. When we relate this constant to the diffusive heat flux in the equilibrium state $\kappa \Delta T/H$ (in dimensional units), we get a definition of the Nusselt number $Nu$ \index{Nusselt number} which is given by 
\begin{equation}
Nu(x_3)=\sqrt{Ra Pr}\langle u_3 T\rangle_{A,t}-\frac{\partial \langle T\rangle_{A,t}}{\partial x_3}=\mbox{const}\,.
\label{eq:Nusselt_Bou}
\end{equation}   
The first term is the convective heat current $J_c$, i.e., heat which is carried by fluid motion from the bottom to the top. The second term is the diffusive heat current $J_d$ which is due to thermal conduction. The \textit{mean thermal boundary layer thickness} is subsequently given by $\delta_T=1/(2 Nu)$ where the factor of 2 captures the approximate (symmetric) temperature drop by $\Delta T/2$ across each boundary layer (top and bottom). An additional average over the vertical direction results in
\begin{equation}
Nu=1+ \sqrt{Ra Pr}\langle u_3 T\rangle_{V,t} \ge 1\,.
\label{eq:Nusselt_Bou1}
\end{equation}   
Note that we did not distinguish our notation between dimensional and dimensionless forms here for simplicity. Equation \eqref{eq:Nusselt_Bou1} contains the diffusive contribution as the constant offset of +1.
\begin{figure*}[t]
\centering
\includegraphics[width=0.9\linewidth]{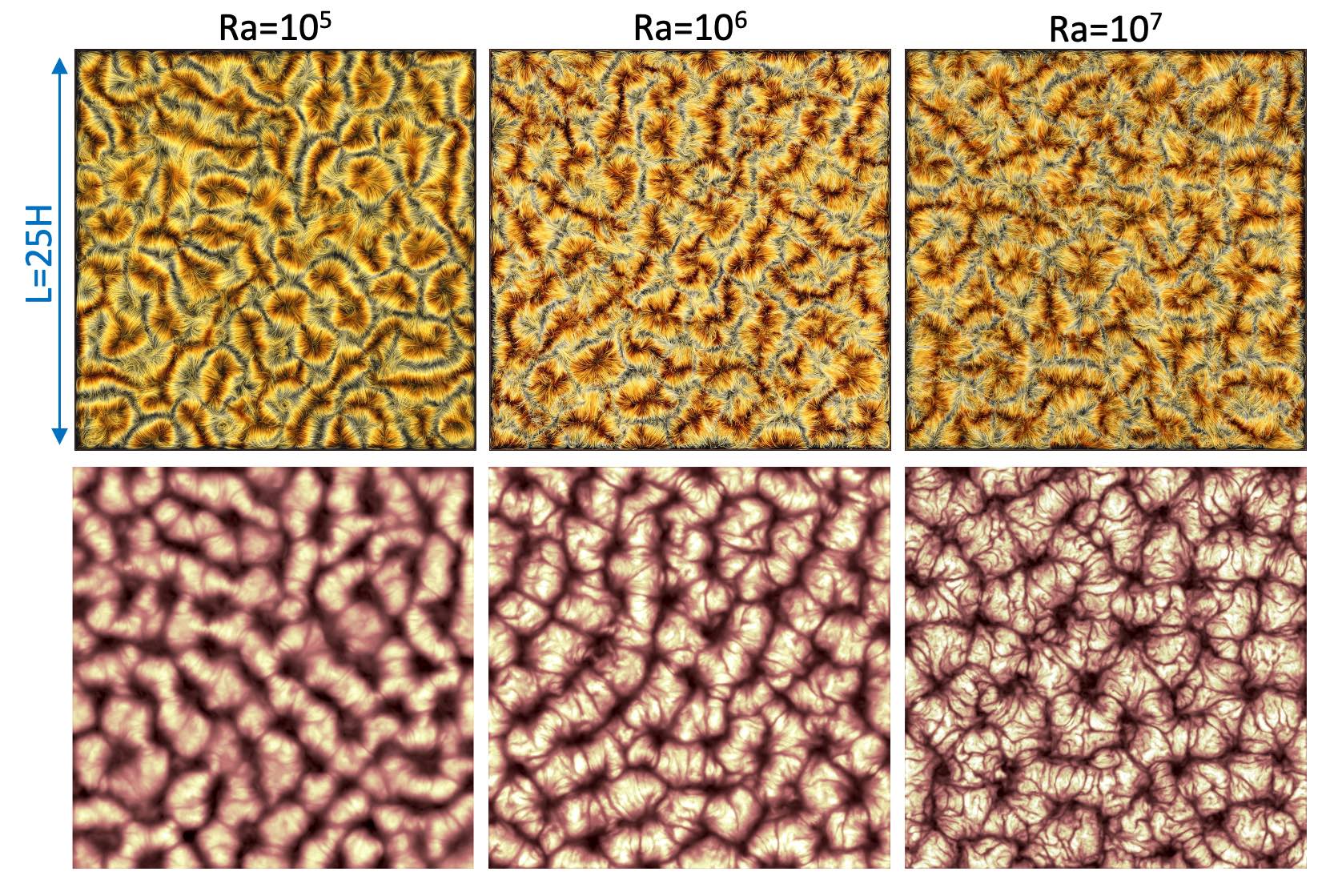}
\caption{Mesoscale simulations at $Pr=0.001$ in a plane convection layer with $\Gamma=L/H=25$ for three different Rayleigh numbers. The top row shows the streamline plots viewed from above. The bottom row displays the corresponding temperature fields inside the thermal boundary layer. All data are taken from ref. \cite{Pandey2022}.} 
\label{fig:pr0001}
\end{figure*}
\begin{figure*}[t]
\centering
\includegraphics[width=0.9\linewidth]{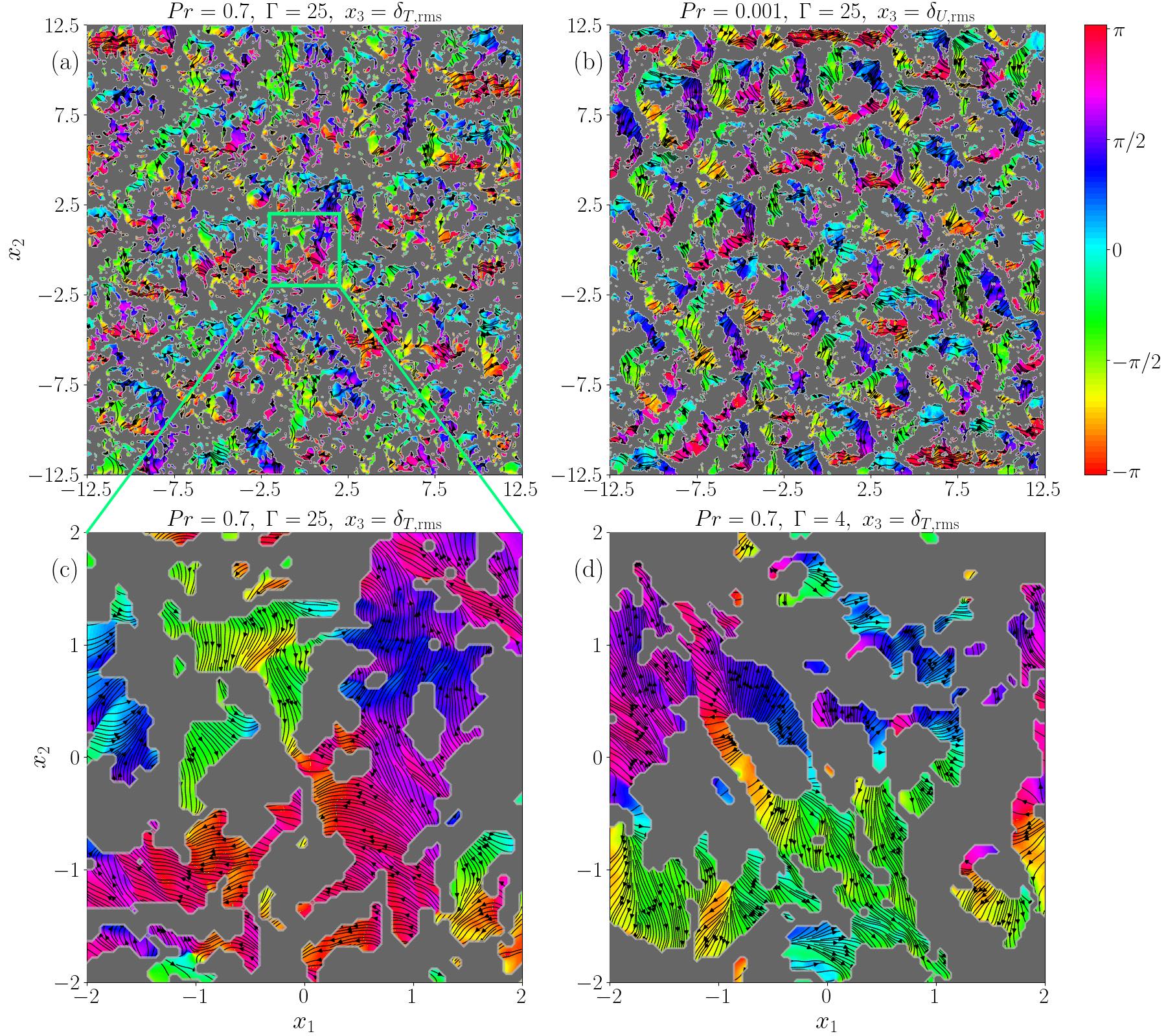}
\caption{Snapshots of coherent shear-dominated and incoherent shear-free regions near the wall for mesoscale convection at $Pr=0.7$ in panels (a,c,d) and $Pr=0.001$ in panel (b). Compared are convection flows in plane layers at $\Gamma=25$ in panels (a,b,c) and $\Gamma=4$ in panel (d). Panel (c) is a zoom of panel (a), see the boxed area in (a). The color bar to the right indicates the local horizontal velocity field orientation in the coherent regions, superposed with corresponding velocity streamlines. Gray regions in all panels stand for incoherent areas. The corresponding area fractions are summarized in table \ref{tab:coh}.} 
\label{fig:coh}
\end{figure*}

\section{Very low Prandtl number}
\label{sec3}
As discussed in the introduction, convection in stellar interiors typically proceeds at very low Prandtl numbers which cannot be obtained in laboratory experiments.  They allow for Prandtl numbers $Pr \gtrsim 0.005$ when opaque liquid metals such as sodium, mercury or gallium-indium-tin alloys are used \cite{Horanyi1999,Glazier1999,Vogt2018,Zuerner2019}. In this section, we discuss DNS of the Boussinesq case at Prandtl numbers below this threshold. 

We conducted DNS for $Pr = 0.001$ and $Ra$ between $10^5$ and $10^7$ in a closed square cuboid of dimensions $(L,L,H)$ with $L = 25H$ (the aspect ratio $\Gamma=L/H=25$) with thermal Dirichlet conditions at the top and bottom, as well as thermally insulated side walls \citep{Pandey2022}. As mentioned above, this Prandtl number cannot be achieved in controlled laboratory experiments. Due to a huge disparity in the kinematic viscosity and thermal diffusivity of the fluid, the convective flows that are generated possess much finer scales in the velocity field than that present in the temperature field. This can be seen Fig.~\ref{fig:pr0001}, which shows the velocity streamlines and the temperature contours in a horizontal plane $A=L^2$ located within the thermal boundary layer region. The view is from the top. Both fields display TSS patterns in the form of rolls which become a bit more cellular with growing $Ra$ due to cross modulations.

It is clear that the velocity streamlines display organization on much finer scales compared to that exhibited by the temperature field (which is subject to strong diffusion). Such fine velocity scales pose a challenge in terms of the spatial resolution that is required to study convective flows for very low Prandtl numbers. Thus, a mesh of size $20480^2 \times 1280$, i.e. having more than half-a-trillion grid points, was used for the highest accessible Rayleigh number of $Ra = 10^7$. These simulations are extremely demanding, even using FDM \cite{Krasnov2011,Pandey2022}, and allowed total simulation times of $\sim 10 \tau_{\rm f}$ only. Here, the free-fall time is given by $\tau_{\rm f}=\sqrt{H/(g\alpha\Delta T)}$ and defines the convective time unit.

\begin{table*}
\begin{center}
\begin{tabular}{lccccccccccc}
\hline
$Ra$  & $Pr$ & $\Gamma$ & $Nu$ & $Re$ & $\delta_T/H$ & $\delta_{T,{\rm rms}}/H$ & $\delta_{U,{\rm rms}}/H$ & $N_{\rm box}$ & $A_{\rm coh} [\%]$ \\
\hline
$10^5$  & $0.7$   & $25$    & $ 4.26 \pm 0.02 $  & $ 92 \pm 0.4 $   & $1.1 \times 10^{-1}$ & $1.2 \times 10^{-1}$ & $1.5 \times 10^{-1}$ & $512^2$ & 41.4 \\
$10^6$  & $0.7$   & $25$    & $ 8.10 \pm 0.03 $  & $ 290 \pm 1 $    & $6.2 \times 10^{-2}$ & $6.1 \times 10^{-2}$ & $1.2 \times 10^{-1}$ & $512^2$ & 40.2 \\
$10^7$  & $0.7$   & $25$    & $ 15.48 \pm 0.06 $ & $ 864 \pm 3 $    & $3.2 \times 10^{-2}$ & $3.1 \times 10^{-2}$ & $1.0 \times 10^{-1}$ & $512^2$ & 39.6 \\
$10^8$  & $0.7$   & $25$    & $ 30.31 \pm 0.05 $ & $ 2510 \pm 7 $   & $1.6 \times 10^{-2}$ & $1.6 \times 10^{-2}$ & $6.7 \times 10^{-2}$ & $512^2$ & 39.7 \\
$10^5$  & $0.7$   & $4$     & $ 4.27 \pm 0.15 $  & $ 93 \pm 2.6 $   & $1.2 \times 10^{-1}$ & $1.3 \times 10^{-1}$ & $2.4 \times 10^{-1}$ & $100^2$ & 40.6 \\
$10^6$  & $0.7$   & $4$     & $ 8.15 \pm 0.19 $  & $ 296 \pm 5 $    & $6.1 \times 10^{-2}$ & $6.1 \times 10^{-2}$ & $1.7 \times 10^{-1}$ & $100^2$ & 40.9 \\
$10^7$  & $0.7$   & $4$     & $ 15.60 \pm 0.90$  & $ 892 \pm 17 $   & $3.2 \times 10^{-2}$ & $3.1 \times 10^{-2}$ & $1.1 \times 10^{-1}$ & $100^2$ & 40.4 \\
$10^8$  & $0.7$   & $4$     & $ 30.40 \pm 1.90$  & $ 2571 \pm 47 $  & $1.6 \times 10^{-2}$ & $1.5 \times 10^{-2}$ & $6.5 \times 10^{-2}$ & $100^2$ & 40.6 \\
$10^5$  & $0.001$ & $25$    & $ 1.21 \pm 0.005 $ & $ 4800 \pm 30 $  & $4.1 \times 10^{-1}$ & $5.0 \times 10^{-1}$ & $1.1 \times 10^{-1}$ & $400^2$ & 41.8 \\
$10^6$  & $0.001$ & $25$    & $ 2.48 \pm 0.005 $ & $ 19876 \pm 1 $  & $2.0 \times 10^{-1}$ & $5.0 \times 10^{-1}$ & $4.6 \times 10^{-2}$ & $400^2$ & 42.4 \\
$10^7$  & $0.001$ & $25$    & $ 4.57 \pm 0.01 $  & $ 56256 \pm 16 $ & $1.1 \times 10^{-1}$ & $2.1 \times 10^{-1}$ & $3.3 \times 10^{-2}$ & $512^2$ & 42.0 \\
\hline
\end{tabular}
\caption{Parameters for the mesoscale convection runs in the Boussinesq limit. We list Rayleigh number $Ra$, Prandtl number $Pr$, aspect ratio $\Gamma$, Nusselt number $Nu$, Reynolds number $Re$, mean thermal boundary layer thickness $\delta_T=H/(2 Nu)$, temperature fluctuation thickness $\delta_{T,{\rm rms}}$, and velocity fluctuation thickness $\delta_{U,{\rm rms}}$. Similar to \cite{Samuel2024}, we determine the distance of the first maxima of temperature and velocity fluctuation profile away from the wall snapshot by snapshot first and take subsequently the arithmetic average to obtain the fluctuation thicknesses $\delta_{T,{\rm rms}}$ and $\delta_{U,{\rm rms}}$, respectively. For low-$Pr$ DNS data, see also \cite{Pandey2022}. Furthermore, we list the number of subvolumes (or square boxes) $A_i$ which cover the cross section plane $A=L^2$ and the coherent near-wall flow area fraction $A_{\rm coh}$.}
\label{tab:coh}
\end{center}
\end{table*}

In ref. \cite{Pandey2022}, we also computed the integral heat and momentum transport, however based on three Rayleigh number values only. It was found that the Nusselt number, the dimensionless measure of turbulent heat transfer, scales as $Nu\sim Ra^{0.29}$ (see eq. \eqref{eq:Nusselt_Bou}) and the Reynolds number, the dimensionless measure of turbulent momentum transfer, as $Re\sim Ra^{0.5}$, see also table \ref{tab:coh}. The Reynolds number is defined as 
\begin{equation}
Re = \sqrt{\frac{Ra}{Pr}} u_{\rm rms} \quad \mbox{with} \quad u_{\rm rms} = \sqrt{\langle {\bm u}^2\rangle_{V,t}}\,,
\label{eq:Reyn}
\end{equation}   
These power laws agree well with those observed in moderate- and high-$Pr$ convection, see e.g. \cite{Bailon2010,Scheel2017,Pandey2018,Stevens2024}, even at different geometries. The highly inertial character of the fluid turbulence in low-Prandtl-number convection is clearly visible when the Reynolds numbers in table \ref{tab:coh} for $Pr=0.001$ are compared to those of $Pr=0.7$ at the same aspect ratio and Rayleigh number. $Re$ is enhanced by a factor of 70.

As the Reynolds number increases with decreasing $Pr$, a wider range of scales in the velocity field is present in low-$Pr$ convection \cite{Schumacher2015}. Further, the integral scale is larger in horizontally-extended domains due to the presence of TSS. This makes extended low-$Pr$ turbulent convection well suited for exploring the energy cascade. We therefore considered the bulk region of the convective flow that is away from the horizontal plates and found that the conditions for the homogeneity and isotropy are fulfilled in this region. We computed the kinetic energy spectrum $E(k)$ from the velocity field in the midplane and found that $E(k)$ exhibits a $k^{-5/3}$ scaling in the inertial range, where $k$ stands for the wavenumber in the midplane. Note that the sum of $E(k)$ over all wavenumbers yields the kinetic energy in the midplane. We also observed the $k^{-5/3}$ scaling for $Pr = 0.005$ and $Pr = 0.021$ flows, and noted that the inertial range becomes wider with increasing $Re$. Therefore, we clearly found that the cascade of the kinetic energy in low-$Pr$ convection is of the Kolmogorov-type~\citep{Verma2018}. 

We also estimated the characteristic spatial scale of TSS for these low $Pr$ which can be considered as a granule size. The scale is denoted as $\Lambda_{\rm TSS}$ and defined as the width of a pair of rolls as observed in Fig.~\ref{fig:pr0001}. We found that the characteristic scale $\Lambda_{\rm TSS} \approx 3H$ for $Ra \le 10^6$; the scale increases to $\Lambda_{\rm TSS} \approx 3.6H$ for $Ra = 10^7$. Furthermore, we also found that $\Lambda_{\rm TSS}$ increases with increasing $Pr$ (at fixed $Ra$) and exhibits a maximum at $Pr \approx 7$. The data exhibits a tendency to saturate at $\Lambda_{\rm TSS} \approx 3H$ when $Pr$ is lowered below 0.02.

How do the velocity field patterns near the wall change when the Prandtl number gets so small? It was shown in \cite{Samuel2024} that the local orientation of the boundary layer flow varies strongly. Thus, a coherent global boundary layer flow is absent, $\langle {\bm u}_h(x_3)\rangle_{A,t}=0$ for $t\gg 1$ with ${\bm u}_h=(u_1,u_2)$. We demonstrated there that the velocity field is rather a collection of local patches of coherent shear-dominated motion, which can be assigned loosely to the circulation roll patterns, i.e. TSS or SG with superposed granules. The coherent flow regions are interspersed by incoherent shear-free regions, the location where the thermal plume clusters rise (or fall) into the bulk. Figure \ref{fig:coh} compares the analysis of the near-wall flow patterns for different Prandtl numbers and aspect ratios. At the edge of the boundary layers, we cover the horizontal plane by $N_{\rm box}$ disjoint square boxes of area content $A_i=A/N_{\rm box}$, where $A=L^2$.  We then calculate the mean horizontal velocity $\bar{\bm u}_h(A_i)$ in each $A_i$ and decompose the cross section into coherent shear-dominated and incoherent shear-free boundary layer regions for $|\bar{\bm u}_h(A_i)| > u^h_{\rm rms}(x_3)$ and $|\bar{\bm u}_h(A_i)|\le u^h_{\rm rms}(x_3)$, respectively. Here, the chosen height of the analysis plane is $x_{3}=\delta_{T,{\rm rms}}$ for $Pr=0.7$ and $x_3=\delta_{U,{\rm rms}}$ for $Pr=0.001$. The temperature fluctuation thickness $\delta_{T,{\rm rms}}$ is smaller than the velocity fluctuation thickness $\delta_{U,{\rm rms}}$ for $Pr=0.7$; the opposite is the case for $Pr=0.001$, see table \ref{tab:coh}. We took the fluctuation thickness, that is closest to the wall, in each case since the dynamics in the vicinity of the bounding walls was in the focus. 

The area fraction of coherent near-wall motion $A_{\rm coh}$ is given in the last column of table \ref{tab:coh}. Independently of the aspect ratio, the Prandtl number, and the Rayleigh number, we see that this area fraction is always approximately 40\% and thus seems to be a very generic feature of MC in this RBC setup. This becomes also obvious when comparing a zoom with an area content of $4H\times 4H$ with the corresponding simulation of \cite{Samuel2024}, cf. panels (c) and (d) of Fig. \ref{fig:coh}. Furthermore, the coherent regions in Fig. \ref{fig:coh} are colored with respect to the local orientation angle of the horizontal velocity $\varphi=\arctan(u_2/u_1) \in [-\pi,\pi]$ to underline the different orientation of the coherent local shear motion. The superposed velocity streamlines underline the different flow orientations of the shear-dominated regions. Gray regions stand for incoherent shear-free areas. It can be concluded that the organization of the near-wall dynamics follows the same principles, relatively independent of aspect ratio $\Gamma$ and $Pr$. This suggests that the structure formation processes at the walls at the top and bottom are dominantly local ones.

\section{Role of boundary conditions}
\label{sec4}

\subsection{Temperature boundary conditions determine large-scale structure formation}
\begin{figure}[t]
\centering
\includegraphics[width=0.95\linewidth]{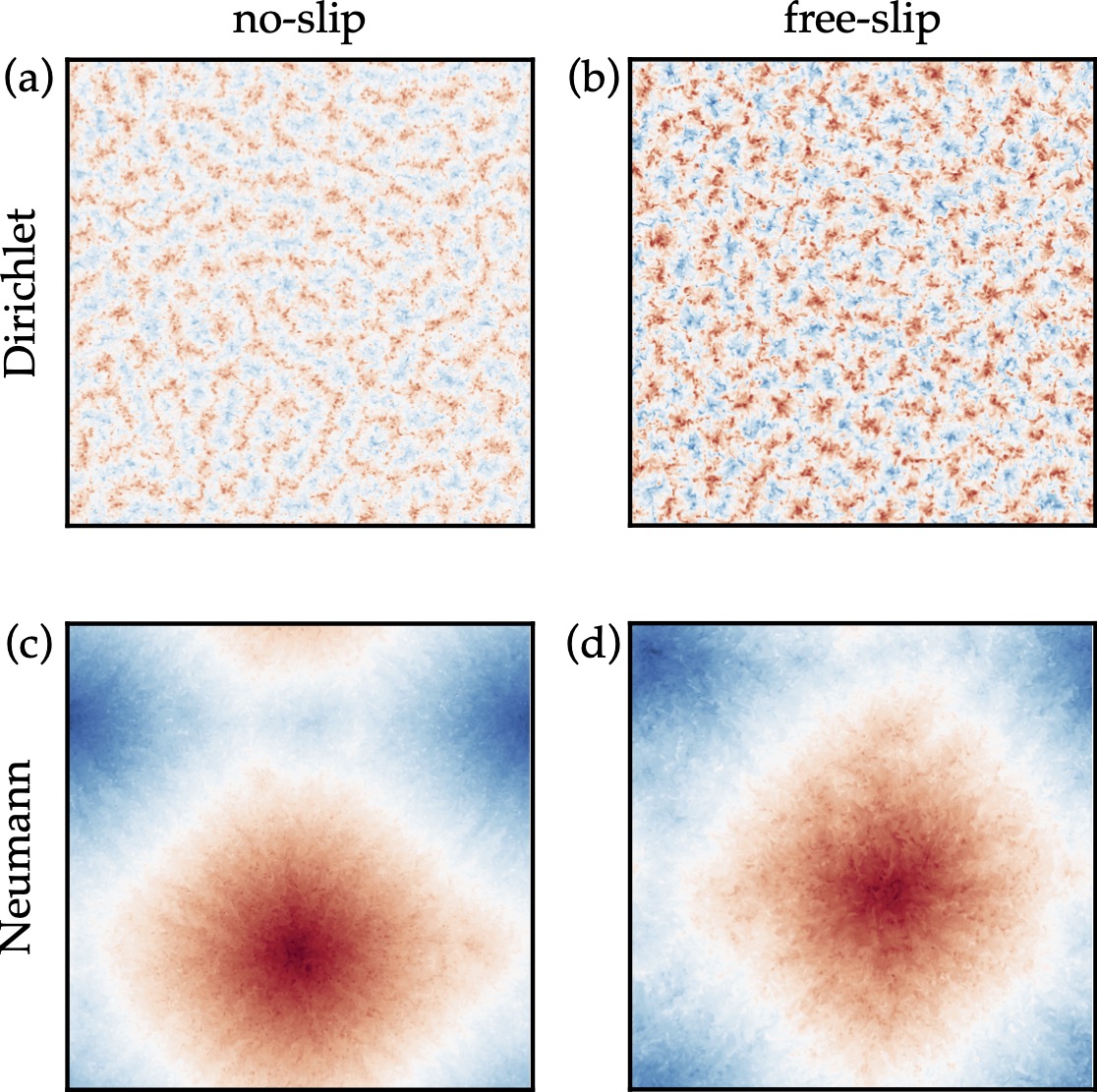}
\caption{
Impact of different mechanical (columns) and thermal (rows) boundary conditions on the formation of long-living large-scale flow structures.  We find the emergence of either (a, b) turbulent superstructures or (c, d) supergranules in the instantaneous temperature field at midplane, $T(x_1,x_2,x_3 = 0.5, t = t_{0})$. $Pr = 1$ and $Ra \approx 10^{6}$ in a horizontally extended and periodic domain of an aspect ratio of $\Gamma = 60$. Data adopted from \cite{Vieweg2021, Vieweg2023a}.} 
\label{fig:mechanical_and_thermal_BCs}
\end{figure}

As discussed in the introduction, turbulent mesoscale convection flows are multi-scale and their dynamical behaviour comprises mechanisms which lead to the manifestation of some long-living large-scale flow structures, the latter of which set themselves apart from much more chaotic turbulence acting on shorter time and length scales. On the one hand, diffusion processes govern the smallest eddies by determining the Kolmogorov (length) scale $\eta_{\textrm{K}} \ll H$ and thus the associated rapid temporal fluctuations. On the other hand, the properties of the long-living large-scale flow structures are governed by both (i) the horizontal extension of the domain as well as (ii) thermal and mechanical boundary conditions. Depending on the particular conditions, their characteristic horizontal extension may be at least $\Lambda_{\textrm{LLFS}} \gtrsim \mathcal{O} \left( H \right)$ together with an associated time scale $\tau_{\textrm{LLFS}} \gg \tau_{\textrm{f}}$. Note that the characteristic temperature unit depends on the thermal boundary conditions: in the Dirichlet case, $T_{\textrm{char}} = \Delta T = T_{\textrm{bot}} - T_{\textrm{top}} > 0$, whereas in the Neumann case, $T_{\textrm{char}} = \beta_{0} H$, cf. eqns. \eqref{Dirichlet} and \eqref{Neumann}. Due to this clear scale separation and larger number of dependencies of the properties of LLFS, which transfer most of the heat across the fluid layer \cite{Krug2020, Vieweg2021,Stevens2024}, the latter represent an active field of research.

If the horizontal extent of the fluid domain is relatively small compared to the size or horizontal extension of the emerging long-living large-scale flow structures, the latter's properties are unavoidably affected by the configuration of the domain. If, in contrast, the domain is significantly larger that the preferred extension of these flow structures, one can expect them to converge with respect to their size to an intrinsically characteristic value. Hence, LLFS manifest differently in different domains. 

When the aspect ratio is small with $\Gamma = L / H \approx 1$, a so-called \textit{large-scale circulation} (LSC) or mean wind throughout the domain forms which has been studied in detail in, e.g., \cite{Ahlers2009}. Its extent $\Lambda_{\textrm{LSC}} \approx H$ is thus (independently of the thermal boundary conditions \cite{Foroozani2021}) clearly influenced or determined by the domain size. Even smaller domains with $\Gamma \ll 1$ may offer more complicated but smaller flow configurations \cite{Zwirner2020,Iyer2020}, whereas larger domains weaken the impact of the lateral boundaries on the flow structures. As the importance of lateral boundaries decreases with $\mathcal{O} \left( \Gamma^{-2} \right)$ \cite{Cross2009, Koschmieder1993, Manneville2006}, one may argue that domains of $\Gamma \gtrsim 16$ are \enquote{large} and more or less close approximations of infinitely extended fluid layers \cite{Koschmieder1993, Stevens2018, Pandey2018, Krug2020}. In terms of turbulent heat and momentum transfer, an independence of the aspect ratio is already found for $\Gamma\gtrsim 4$ in ref. \cite{Stevens2024} when Dirichlet boundary conditions are applied for the temperature field. 

As will become clearer below, it is in fact the ratio between the size of the large-scale flow structures and the domain size, $\Lambda_{\textrm{LLFSs}} / H$, which determines the significance of lateral boundaries. When this ratio increases, the latter's impact decreases and dependencies on other boundary conditions govern the nature of the long-living large-scale flow structures. As discovered recently \cite{Vieweg2021} and highlighted in Fig. \ref{fig:mechanical_and_thermal_BCs} for a very large aspect ratio domain of $\Gamma = 60$, thermal boundary conditions determine the pattern formation, whereas the impact of mechanical boundary conditions on this self-organization of the flow is much weaker and becomes subdominant. In detail, we compared no-slip and free-slip mechanical boundary conditions, cf. eqns. \eqref{eq:noslip} and \eqref{eq:freeslip}. Fixed heat flux conditions lead also in other convection configurations to large-scale structures or shear, such as in a 2D RBC setup without walls \cite{Liu2024}.  

In the traditional Dirichlet case of constant \textit{temperatures} at the plates, cf. eq. \eqref{Dirichlet} or panels (a, b) of Fig. \ref{fig:mechanical_and_thermal_BCs}, one observes the formation of TSS with a time-independent characteristic horizontal extension of roughly $\Lambda_{\textrm{TSS}} \approx 5 H \sim \mathcal{O} \left( H \right)$ \cite{Stevens2018,Pandey2018,Vieweg2021} for $Pr\sim 1$. This self-organization of the flow is fundamentally changed for thermal Neumann boundary conditions of a constant (vertical) \textit{temperature gradient} or heat flux at the plates \cite{Vieweg2021}, see again eq. \eqref{Neumann} and also panels (c, d) of Fig. \ref{fig:mechanical_and_thermal_BCs}. Here, a gradual, time-dependent aggregation eventually leads to a SG \cite{Vieweg2021, Vieweg2022, Vieweg2023, Vieweg2023a, Vieweg2024, Vieweg2024a} with a final size of $\Lambda_{\textrm{SG}} = \Gamma H$. This implies that the final structure is limited in horizontal extension by the domain size only. Although all of these long-living large-scale flow structures are most prominent in the scalar temperature field, they are also present in the velocity field \cite{Pandey2018, Vieweg2021}. Fodor et al. \cite{Fodor2019} detected for flux-driven RBC similar characteristic scales as for the Dirichlet case, which underlines that the supergranule formation with $\Lambda_{\textrm{SG}} \gg \Lambda_{\rm TSS}$ is a very slow dynamical process.  

As LLFSs have also been successfully detected based on Lagrangian tracer trajectory data \cite{Vieweg2021a, Schneide2022, Vieweg2024}, it is possible to quantify their role in heat transfer across the fluid layer based on the finite-time coherence of (potentially sparse) trajectory fragments. In the Dirichlet case, it was clearly shown that regions between the centers of the superstructure rolls contribute most to the turbulent heat transfer in comparison to the almost invariant spatial regions inside the convection rolls.

\begin{figure}[t]
\centering
\includegraphics[width=0.95\linewidth]{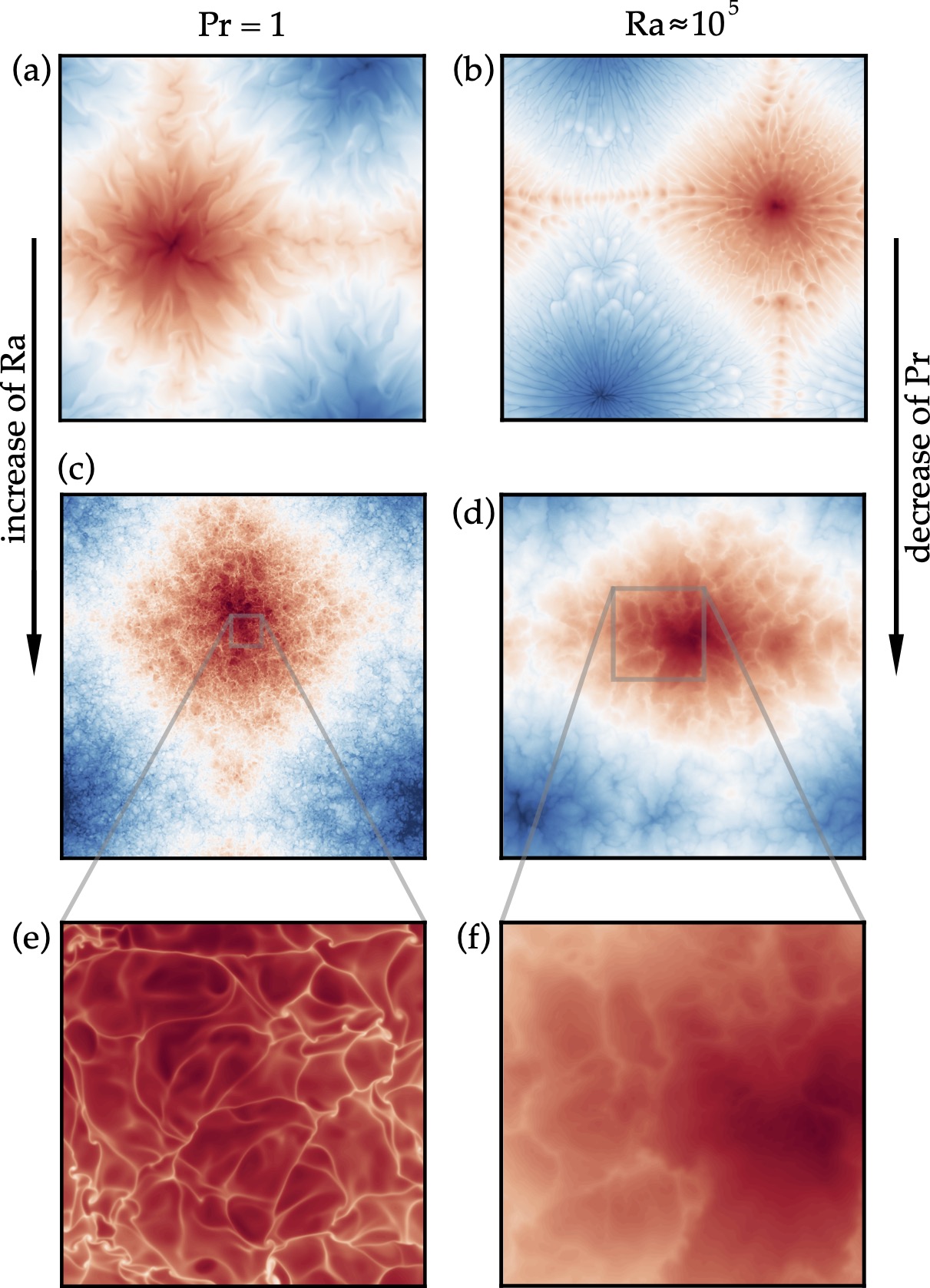}
\caption{Ubiquity of supergranule formation across the numerically accessible parameter space. We find the emergence of the supergranules at $Pr = 1$ from (a) $Ra_{\rm N} \approx 10^{4}$ to (c) $Ra_{\rm N} \approx 10^{8}$ as well as at $Ra_{\rm N} \approx 10^{5}$ from (b) $Pr = 10^{2}$ to (d) $Pr = 10^{-2}$. As shown via the magnifications in panels (e, f), the supergranules are superposed to granule structures. The prominence of the latter varies with $Pr$. Here we apply free-slip boundary conditions in a horizontally periodic domain of $\Gamma = 60$ and visualise instantaneous $T(x_1,x_2,x_3=1-\delta_T/2,t = t_{0})$.  Data is adopted from \cite{Vieweg2021, Vieweg2023a, Vieweg2024a}.}
\label{fig:ubiquity_of_SGA}
\end{figure}
\begin{figure*}[t]
\centering
\includegraphics[width=0.85\linewidth]{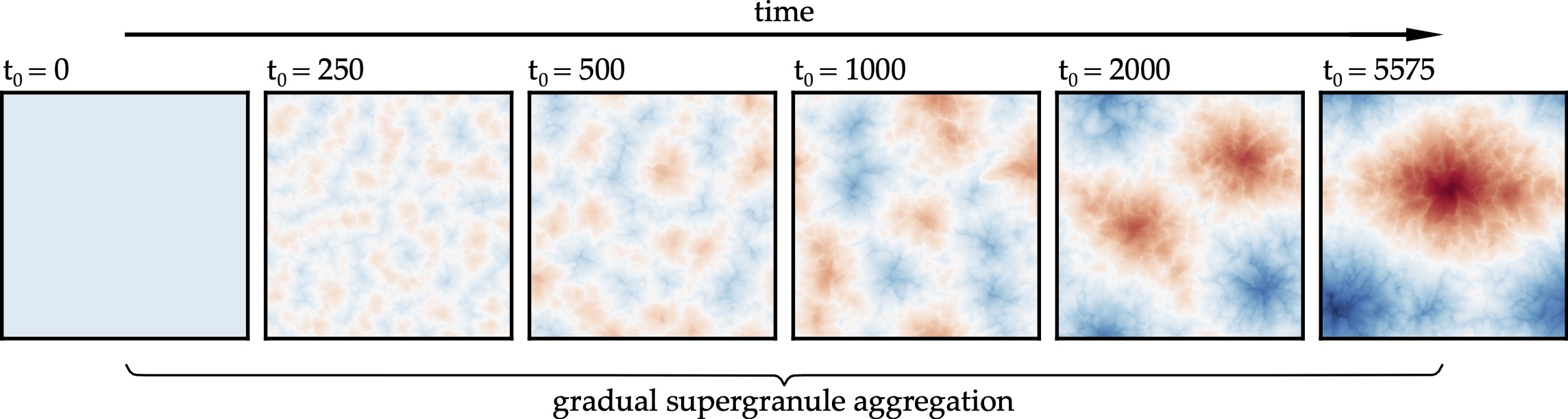}
\caption{Transient gradual supergranule aggregation. The supergranule is the eventual result of a transient gradual aggregation process. The duration of this aggregation depends on both $Ra$ \cite{Vieweg2021} and $Pr$ \cite{Vieweg2024a} and may take $\mathcal{O} \left( 10^{4} \tau_{\textrm{f}} \right)$. Despite the strong turbulence with $\textrm{Re} \approx 2000$ at $Ra_{\rm N} \approx 10^{5}$ and $Pr = 10^{-2}$, this process appears to be driven by secondary instabilities \cite{Chapman1980, Chapman1980a, Vieweg2021}. Here we apply free-slip boundary conditions in a horizontally periodic domain of $\Gamma = 60$ and visualize instantaneous $T(x_1,x_2, x_3=1-\delta_T/2, t = t_{0})$. Note that the right-most panel is identical to Fig. \ref{fig:ubiquity_of_SGA}(d). Data is adopted from \cite{Vieweg2023a, Vieweg2024a}.} 
\label{fig:evolution_gradual_supergranule_aggregation}
\end{figure*}

Interestingly, SGs seem to be a ubiquitous feature of heat flux-driven convection. They have been found for both $10^{4} \lesssim Ra_{\rm N} \lesssim 10^{8}$ (given $Pr = 1$) \cite{Vieweg2021} as well as $10^{-2} \leq Pr \leq 10^{2}$ (given $Ra_{\rm N} \equiv Ra Nu \approx 10^{5}$) \cite{Vieweg2024a}, see also Fig. \ref{fig:ubiquity_of_SGA}. Of course, larger $Ra$ and smaller $Pr$ are of interest in the geophysical and astrophysical context. However, they are neither numerically accessible for such horizontally strongly extended domains, nor is there evidence that the supergranules vanish beyond certain critical $Ra$ or $Pr$.  Note further that supergranules are superposed to smaller (yet large-scale) sub-dominant flow structures of roughly $\Lambda_{\textrm{G}} \approx 4 H$ \cite{Vieweg2021, Vieweg2023a, Vieweg2024} which are termed \textit{granules} (G) and highlighted in Fig. \ref{fig:ubiquity_of_SGA}(e). It is interesting to observe this hierarchy of structures, which we introduced in section 1 for solar convection, in the strongly simplified RBC configuration. 

The emergence of such fundamentally different characteristic flow structures in the Neumann case is in accordance with the changed character of the primary linear instability of the convection layer for these different boundary conditions. While the critical wavelength at the onset of convection with Dirichlet conditions is $\lambda_{\textrm{crit}} = 2\sqrt{2}$ and $\approx 2.02$ for free-slip \cite{Rayleigh1916} and no-slip conditions \cite{Pellew1940,Chandrasekhar1961}, respectively, the critical wavelength is $\lambda_{\textrm{crit}} = \infty$ \cite{Hurle1967} for the Neumann case, independently of the mechanical boundary conditions \cite{Vieweg2022}. As shown via the time series in Fig. \ref{fig:evolution_gradual_supergranule_aggregation}, the supergranule is (i) the result of a long transient gradual supergranule aggregation and (ii) affected by the specific domain configuration only at late times when $\Lambda_{\textrm{SG}} / \Gamma \rightarrow 1$. 

A numerical leading Lyapunov vector stability analysis \cite{Pikovsky2016} of the fully turbulent flow has revealed that the time-dependent, three-dimensional temperature perturbation field $\delta T({\bm x},t)$ exhibits strong instabilities at the (time-dependent) transient supergranule scale, thus driving the growth of the supergranules until the domain size is reached \cite{Vieweg2021}. Interestingly, this resembles secondary instabilities, which have been known already from analytical studies slightly above the onset of heat flux-driven convection \cite{Chapman1980, Chapman1980a}, and suggests that secondary instabilities survive far into the turbulent regime \cite{Vieweg2021, Vieweg2023a}. On the other hand, an analysis of the three-dimensional spectral energy transfer proved that this transient aggregation process is driven by an inverse cascade of thermal variance within the subset of purely two-dimensional triads or spectral mode interactions \cite{Vieweg2022, Vieweg2023a}. Inverse cascades at large scales have also been found for the kinetic energy \cite{Vieweg2022}. 

So far, only the inclusion of additional physical mechanisms, such as weak rotation around the vertical axis, has allowed to stop the gradual supergranule aggregation at intermediate scales before reaching the domain size. This impact of rotation is also relevant for the structure formation in solar convection, as discussed by Vasil et al. \cite{Vasil2021}. Rotation requires to add a Coriolis term to the left hand side of \eqref{eq:mom_bou},
\begin{equation}
({\bm u}\cdot{\bm \nabla}){\bm u} \to ({\bm u}\cdot{\bm \nabla}){\bm u} + \frac{1}{Ro}({\bm e}_3\times {\bm u})\,
\end{equation}
with a new dimensionless parameter, the Rossby number $Ro$ which is given by 
\begin{equation}
Ro = \frac{\sqrt{\alpha g T_{\textrm{char}}}}{ 2 \Omega \sqrt{H}} \gtrsim 5
\end{equation}
which is a weak rotation \cite{Vieweg2022, Vieweg2023a}. Remarkably, while the self-organisation of the flow is strongly impacted at supergranule scale (but not at granule scale), the overall heat transfer across the fluid layer is barely altered \cite{Vieweg2022, Vieweg2023a}. We point at this place to similar large-scale structure formation processes by inverse cascades in stably stratified rotating thin-layer turbulence without boundary layers that have been studied by large-scale DNS in ref. \cite{Alexakis2024}.
\begin{figure*}[t]
\centering
\includegraphics[width=0.85\linewidth]{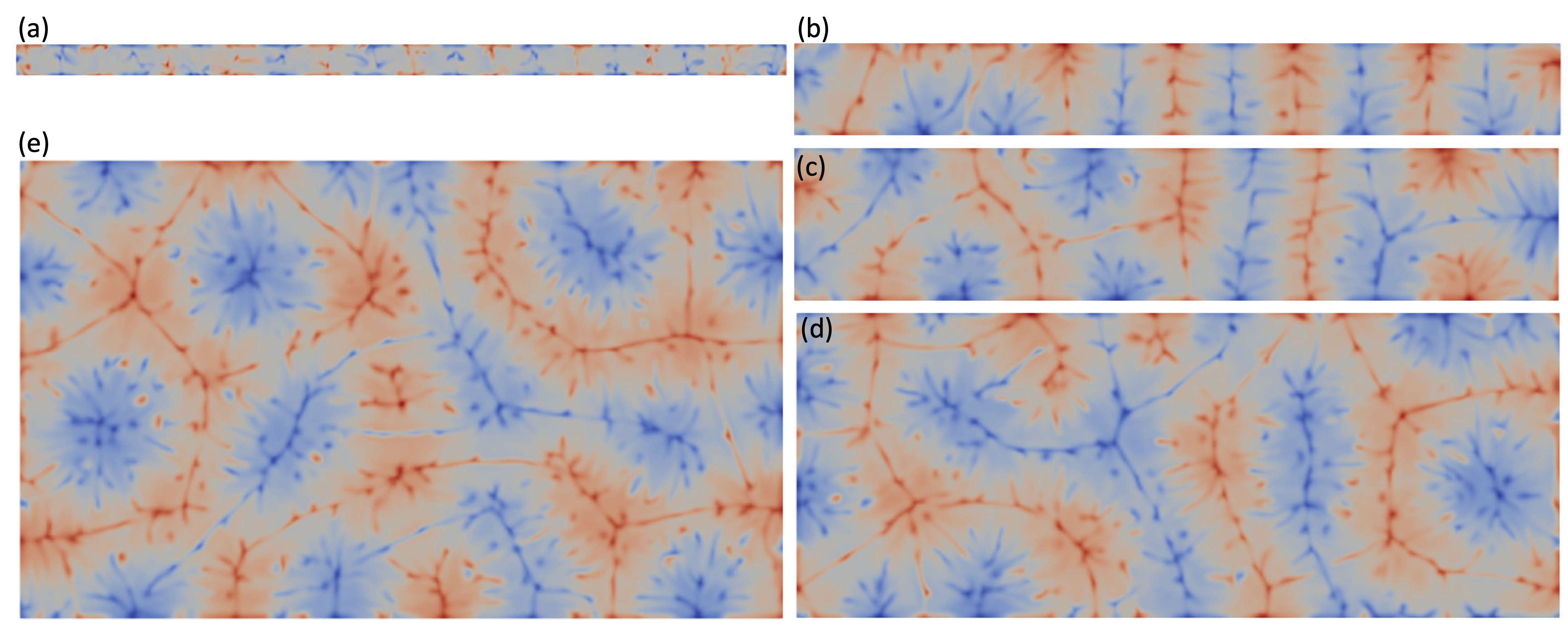}
\caption{Turbulent superstructures in different rectangular cuboids of $\Gamma_1 = 25$ for $Pr = 7$ and $Ra = 10^5$. (a) $\Gamma_2 = 1$, (b) $\Gamma_2 = 3$, (c) $\Gamma_2 = 5$, (d) $\Gamma_2 = 10$, (e) $\Gamma_2 = 15$. The temperature field in mid-horizontal $x_1$--$x_2$ plane is displayed. View is always from the top. The flow organization and the characteristic scale is affected when the aspect ratio $\Gamma_2 \leq 5$.} 
\label{fig:tss_rect}
\end{figure*}
\begin{figure*}[t]
\centering
\includegraphics[width=0.92\linewidth]{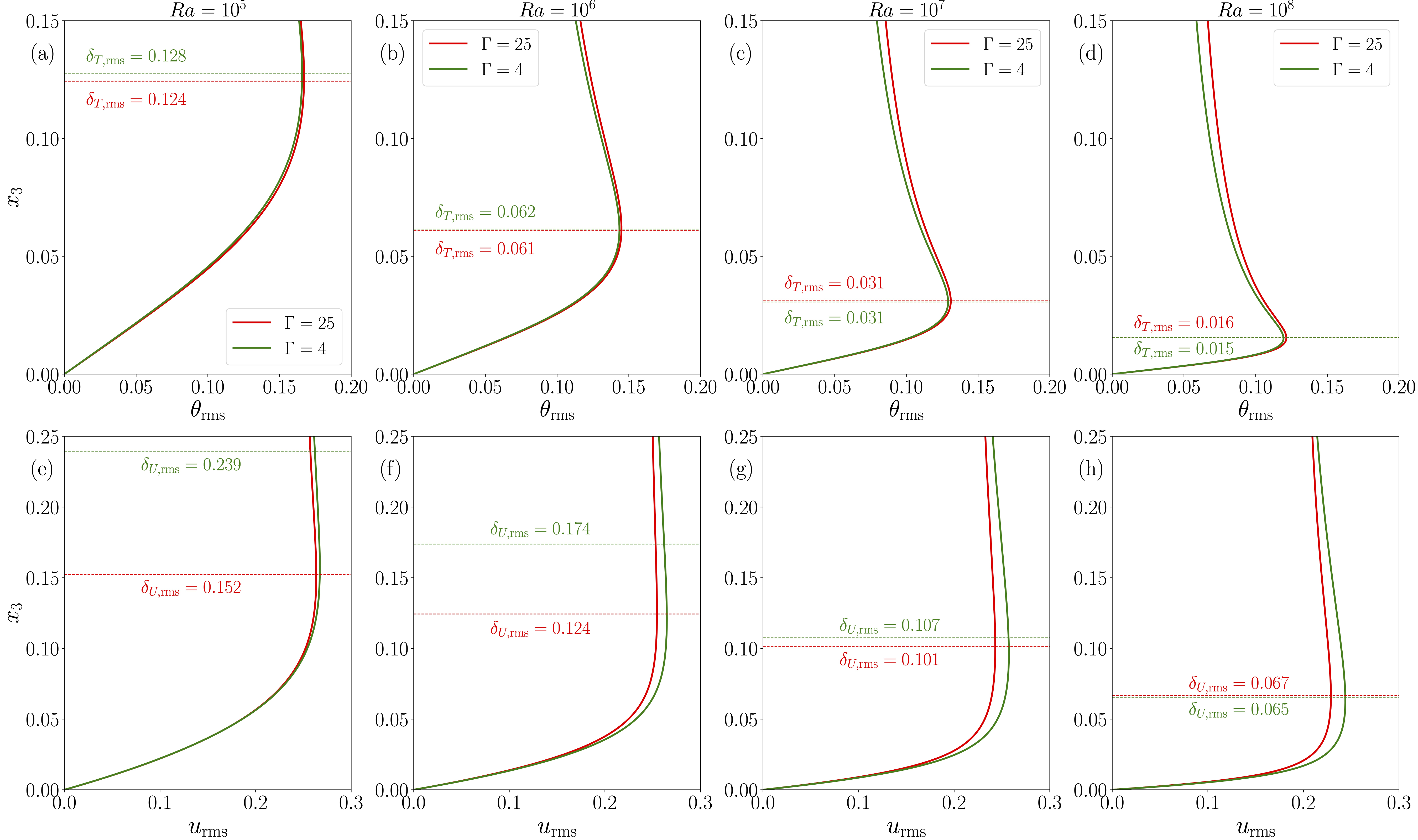}
\caption{Mean fluctuation profiles of temperature (a,b,c,d) and velocity (e,f,g,h) for different aspect ratios (see legend) and Rayleigh numbers (see title). The corresponding fluctuation thicknesses are indicated by value (in units of $H$) and corresponding horizontal lines. Note that $\theta({\bm x},t)=T({\bm x},t)-\langle T(x_3)\rangle_{A,t}$ is taken to get $\theta_{\rm rms}(x_3)$.} 
\label{fig:BLprofile}
\end{figure*}

These studies with idealized thermal boundary conditions provide helpful general guidance for the interpretation of natural flows and laboratory experiments. However, understanding the remaining discrepancies between data from numerical and experimental approaches requires an even deeper understanding of experimentally present boundary and measurement conditions. A digital twin \cite{Vieweg2024b} of one specific laboratory experiment \cite{Moller2022} has recently proven effective in tracing back these discrepancies to their root causes. This included modeling the (to the fluid layer) adjacent solid plates as well as their pressure-driven cooling flow in a direct numerical simulation, making thus use of both conjugate heat transfer and Newton cooling conditions as described by equations \eqref{conj} and \eqref{eq:BC_Newton_cooling}, respectively. This digital twin and its successive simplification towards the classical plate-less Dirichlet conditions \cite{Vieweg2024b} provide an important first step towards more realistic thermal conditions.

The broad variety of recent numerical studies allows to draw the following overall picture: 
\begin{enumerate}
\item Horizontally extended domains are required to allow for a self-organization of the flow based on intrinsic principles, i.e. unaffected by the lateral boundaries. Our studies suggest $\Gamma \gtrsim 16$ as an appropriate domain size, at least for Dirichlet conditions of the temperature field. 
\item Long-living large-scale flow structures, i.e. both TSSs as well as SGs, account for the majority of the heat transfer across the fluid layer. Consequently, these structures determine the scaling laws of $Nu \sim Ra^{\gamma}$. Interestingly, the overall heat transfer seems not to be affected too strongly by the final size of LLFSs at first glance. 
\end{enumerate}
However, there is little but increasing evidence that thermal boundary conditions which deviate from the classical thermal Dirichlet conditions result in an increased heat transfer across the fluid layer. Such thermal conditions are in fact very relevant for geophysical and astrophysical systems offering mesoscale convection and will thus certainly need to be addressed by upcoming numerical studies.

\begin{figure*}[t]
\centering
\includegraphics[width=0.8\linewidth]{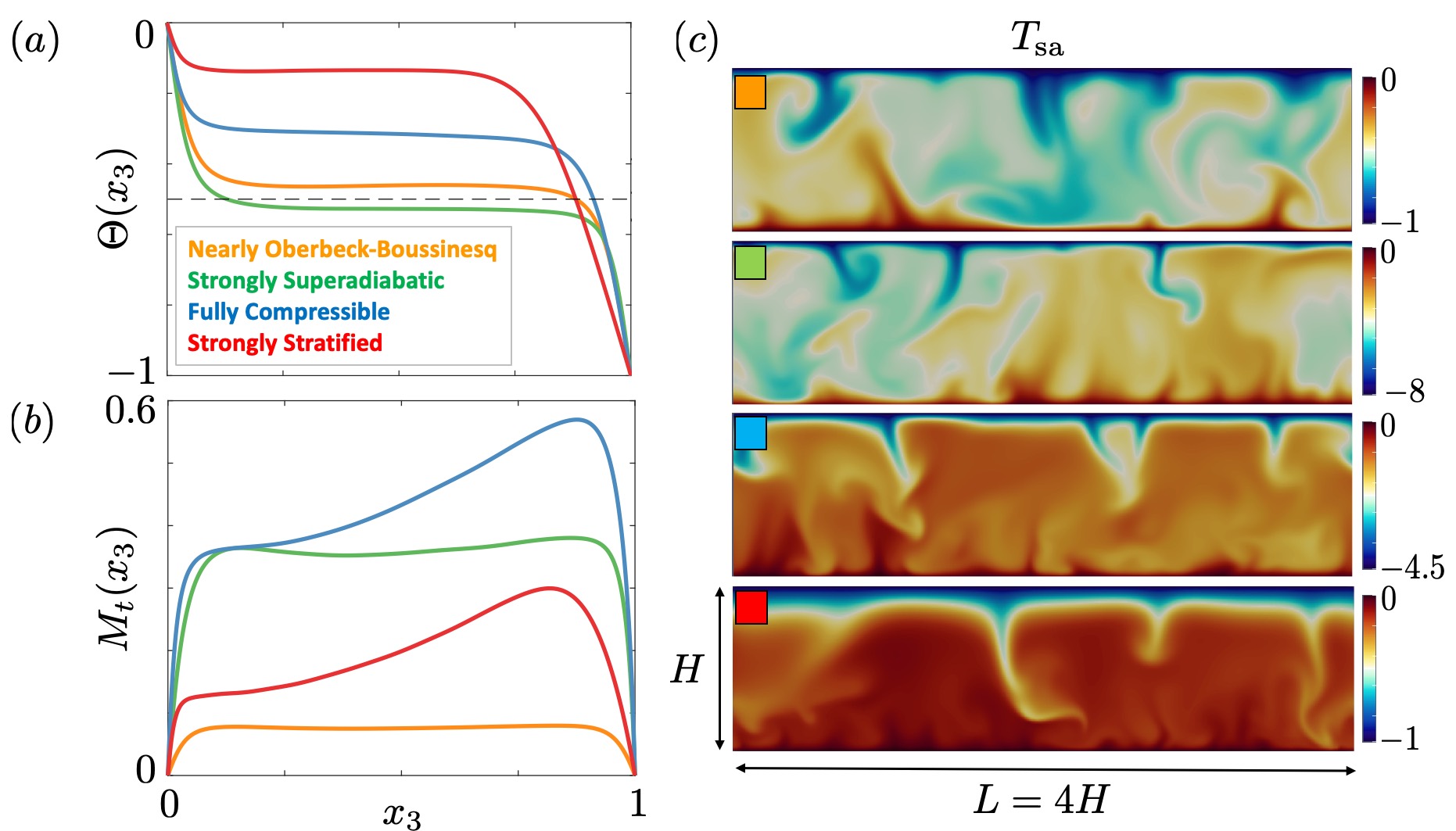}
\caption{Four different regimes of compressible convection. These are the nearly Oberbeck-Boussinesq (OB), the strongly superadiabatic (SAC), the fully compressible (FCC), and the strongly stratified (SSC) convection regime. (a) Vertical profiles of the rescaled mean superadiabatic temperature $\Theta(x_3)$, cf. eq. \eqref{eq:Tsa1}. (b) Vertical profile of the turbulent Mach number $M_t(x_3)$. (c) Four snapshots of the superadiabatic temperature $T_{\rm sa}$. Vertical cross sections are displayed. The color coding is given by the legend in (a). The Rayleigh number is $Ra=10^6$ in all cases. The four DNS are run at $(\varepsilon,D)=(0.1,0.1)$, (0.8,0.1), (0.45, 0.5), and (0.1,0.8) for OB, SAC, FCC, and SSC, respectively. The range of $T_{\rm sa}$ is indicated to the right of each panel.} 
\label{fig:comp}
\end{figure*}

\subsection{Shape of the plane layer} 

Keeping the impact of thermal boundary conditions on the formation and nature of LLFSs in mind, exploring TSSs with applied Dirichlet boundary conditions at high Rayleigh numbers is a challenge as the required computing resources are proportional to the cross-sectional area of the convective layer. To mitigate this restriction, while achieving high turbulence levels in the flow, we are systematically exploring the properties of turbulent superstructures in rectangular cuboids of dimensions $(L_1, L_2, H)$ with $L_1 \geq L_2 \ge H$ in the following subsection. Note that past studies exploring the properties of TSS have utilized domains that are either square cuboids with $L_1 = L_2 = L\gg H$~\citep{Hartlep2003,Hartlep2005,Pandey2018,Stevens2018,Pandey2022,Moller2022} or cylinders with $D \gg H$~\citep{Emran2015,Sakievich2016,Stevens2024} where $D$ is the diameter. It will be helpful if similar superstructure patterns can be realized in rectangular cuboids with $L_2 < L_1$ as this leads to a reduction in the required computing resources. Consequently, the statistical properties and structures of convection can be studied at even stronger thermal forcing (or higher $Ra$) and their dynamical evolution could be probed for longer time scales. Open is how the expected turbulent superstructures will experience the effects of boundaries when $L_2$ approaches $H$ and how this affects the turbulent heat transfer. 

We have thus conducted a series of DNS of  convection in rectangular cuboids of a fixed $\Gamma_1 = L_1/H = 25$ and varying $\Gamma_2 = L_2/H = 1, 3, 5, 10, 15$, and $25$. Figure~\ref{fig:tss_rect} shows the instantaneous temperature field in the midplane of these cuboids for fixed $Pr = 7$ and  $Ra = 10^5$. We observe that the superstructure patterns that are realized in $\Gamma_2 = 10$, and $\Gamma_2 = 15$ domains are mostly similar to those observed in $\Gamma_2 = 25$ domain \cite{Pandey2018}. It can be seen that polygonal cells, with cold fluid sinking in the central region and hot fluid rising through the periphery of these regions, are formed in the cuboids for $\Gamma_2 \geq 10$. However, as the domain is confined further along the horizontal $y$-direction the flow pattern starts to be affected by the boundaries. Even though the flow in $\Gamma_2 = 5$ domain has a tendency to exhibit cellular structures, their formation is not supported any further. Domains with $\Gamma_2 = 3$ and $\Gamma_2 = 1$ do not display cellular structures at all. Further, we observe that the characteristic scale of the superstructure pattern $\Lambda_{\rm TSS}$ remains nearly unchanged when $\Gamma_2$ is decreased from 25 to 5, but starts to decrease with $\Gamma_2$ when $\Gamma_2 \leq 5$. We also compute the global heat and momentum transports and find that the latter decreases systematically with decreasing $\Gamma_2$. The heat transport, however, exhibits a non-monotonic variation which requires further investigation in the future. 

Figure \ref{fig:BLprofile} displays mean fluctuation profiles for the case of $Pr=0.7$, related to the data analysis in Fig. \ref{fig:coh} and table \ref{tab:coh}. The comparison of the data at $\Gamma=4$ and 25 confirms a fair agreement of the temperature and velocity fluctuation profiles. The corresponding distances from the wall at which the profile becomes maximal, the fluctuation thicknesses which are shown as horizontal lines, agree well for the temperature. The differ for the velocity field for $Ra=10^5$, but converge to each other for $Ra=10^8$, see also the detailed values in table \ref{tab:coh}. The table shows also that the global transport measures, $Nu$ and $Re$, agree very well for both aspect ratios which confirms a recent analysis of Stevens et al. \cite{Stevens2024}. We mention here in addition that the horizontal sidewalls for $\Gamma=25$ are thermally insulating and solid while being periodic in case of $\Gamma=4$. Table \ref{tab:coh} also underlines that the mean thermal boundary layer thickness agrees very well with the temperature fluctuation thickness, $\delta_T\approx \delta_{T,{\rm rms}}$.

\section{Role of stratification and temperature-dependent material properties}
\label{sec5}

\subsection{Regimes of compressible convection}
Non-Boussinesq convection can be obtained in two different routes: 
\begin{enumerate}
\item The first one is followed by switching from shallow convection to systems where the scale height, the scale for which thermodynamic state variables (pressure, temperature, density) vary by an order of magnitude, is smaller than the height $H$ of the layer, $H \lesssim {\cal H}_{\beta}$ with $\beta=\{p,T,\rho\}$ \cite{Froehlich1992,Verhoeven2015}. See also eq. \eqref{scaleheight}. This is connected to \textit{compressibility effects} that have to be incorporated. 
\item The second one is followed by including \textit{temperature and pressure dependencies for material parameters}, such as thermal conductivity and dynamical viscosity. This causes already non-Boussinesq effects for a standard RBC setup, as shown in \cite{Horn2013,Pandey2021,Pandey2021a}. 
\end{enumerate}
One essential manifestation of non-Boussinesq effects in convection is when the up-down symmetry with respect to the half-height is broken and $T_c\ne (T_{\rm top}+T_{\rm bot})/2$. In many natural convection processes both routes, that we listed above, appear often as a superposition. We mention here that technological applications of convection are mostly affected by the second route \cite{Valori2019}. In the following, we briefly discuss both routes separately. 

As explained in subsection 2.4, the two additional dimensionless parameters in compressible thermal convection, the superadiabaticity $\varepsilon$ and the dissipation number $D$, are not independent of each other and span a triangular parameter plane that is constrained by $D\le 1-\varepsilon$ on the one hand and by $0\le \varepsilon, D\le 1$ on the other hand, cf. Fig \ref{fig:parameter_triangle}. Different regimes of compressible convection have been identified and investigated in refs. \cite{Panickacheril2023,Panickacheril2023a} for $Pr\sim 1$. These are:
\begin{enumerate}
   \item \textit{Nearly Oberbeck-Boussinesq convection}, which exists for $\varepsilon\ll 1$ and $D\ll 1$ close to the exact OB Rayleigh-B\'{e}rnard convection limit at $(\varepsilon,D)=(0,0)$. This regime will be also denoted to as OB for the following. The limits of both parameters have to be taken as follows: first $\varepsilon\to 0$ which includes the anelastic convection case with $\nabla\cdot(\bar{\rho}{\bm u})=0$. The anelastic limit is not further detailed here, see e.g. refs \cite{Verhoeven2015,Alboussiere2017,Jones2022} and \cite{Curbelo2019} for $Pr\to\infty$. Subsequently the limit $D\to 0$ is taken to get to the OB case.
   \item \textit{Strongly stratified convection} exists for $\varepsilon\ll 1$ and $D\to 1$ and will be denoted as SSC. 
   \item \textit{Fully compressible convection} exists for $\epsilon\approx D \approx 0.5$ and will be denoted as FCC. It is the range in which the free-fall Mach number $M_{\rm f}=U_{\rm f}/c_s$ is maximum. Here $U_{\rm f}=\sqrt{\varepsilon g H}$ is the free-fall velocity and $c_s=\sqrt{\gamma R T_{\rm bot}}$ the speed of sound, see also eq. \eqref{sound}.
   \item \textit{Strongly superadiabatic convection} exists for $D\ll 1$ and $\varepsilon\to 1$ and will be abbreviated by SAC in the following. This regime has been studied in detail in \cite{Panickacheril2023a}. It might be particularly relevant for the surface convection at the Sun \cite{Schumacher2020} even though we neglect radiative transfer and very low $Pr$ for the following considerations \cite{Nordlund2009}.
\end{enumerate}
Figure \ref{fig:comp} compares the four different regimes for a fixed Rayleigh number of $Ra=10^6$ and a Prandtl number $Pr=0.7$. We plot the mean vertical profiles of the superadiabatic temperature field, which is given by  
\begin{equation}
T_{\rm sa}({\bm x},t)=T({\bm x},t)-\bar{T}(x_3)\,,
\label{eq:Tsa}
\end{equation}
see again eq. \eqref{eq:T_ad} in subsection 2.3. For this comparison, we display rescaled temperature profiles, which are given by 
\begin{equation}
\Theta(x_3)=\frac{\langle T_{\rm sa}(x_3)\rangle_{A,t}}{[\langle T_{\rm sa}(0)\rangle_{A,t}-\langle T_{\rm sa}(1)\rangle_{A,t}]}\,.
\label{eq:Tsa1}
\end{equation}
It is seen that particularly, the runs for $D\gg 0$, i.e., FCC and SSC, show a strong offset from the symmetric mean at -0.5 for the profiles in the bulk. In panel (b), we add the turbulent Mach number which is given by
\begin{equation}
M_t(x_3)=\frac{u_{\rm rms}(x_3)}{\gamma R\langle T(x_3)\rangle_{A,t}}\,.
\label{eq:Mt}
\end{equation}
The profiles for FCC and SSC are highly asymmetric across the convection layer, and the highest turbulent Mach numbers are indeed obtained for the FCC case, close to the top boundary layer. Typical snapshots of the superadiabatic temperature field $T_{\rm sa}$ are given in Fig. \ref{fig:comp}(c). Even though the Rayleigh numbers are still small, we found that the thermal BLs at the top and bottom of the layer are highly asymmetric, in particular for the highest $D$ values of the dissipation number $D\gtrsim 0.65$ in the SSC case. For this limit regime of compressible convection, we identified a top boundary, which is mostly stably stratified, interspersed by localized region of detachment of thermal plumes which can fall deep into the highly stratified bulk region of the layer, focused by compressibility \cite{Panickacheril2023}. In this regime, density fluctuations are mostly aligned with pressure fluctuations, and not with those of temperature, as being the case for smaller $D$ values.
\begin{figure*}[ht!]
	\centering
	\includegraphics[width=0.88\linewidth]{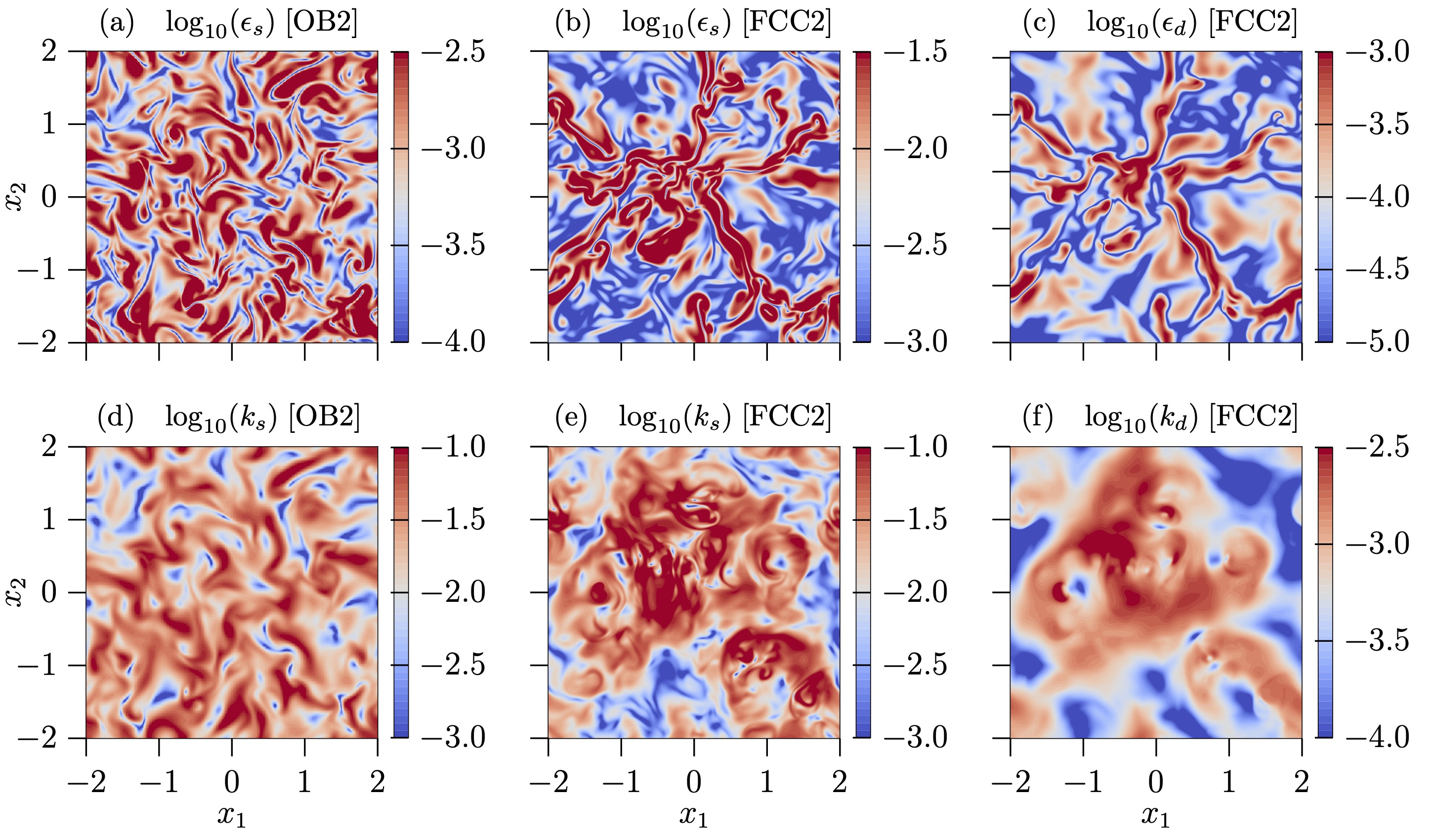}
	\caption{Snapshot contour plots of different components of the turbulent kinetic energy and kinetic dissipation rate fields taken all at half height $x_3=H/2$ of the layer for convection flows at $Ra=10^6$ and $Pr=0.7$. (a) Solenoidal dissipation rate field $\epsilon_s$ in the Oberbeck-Boussinesq (OB) limit. (b) Solenoidal dissipation rate field $\epsilon_s$ in the fully compressible case (FCC) at $\varepsilon=0.45$ and $D=0.5$. (c) Dilatational dissipation rate field $\epsilon_d$ in FCC case. (d) Solenoidal turbulent kinetic energy $k_s$ in (OB) limit. (b) Solenoidal kinetic energy $k_s$ in FCC case. (c) Dilatational kinetic energy $k_d$ in FCC case. All positive definite fields are plotted in units of the decadic logarithm. The corresponding color bar displays the range of amplitudes.  Kinetic energy and kinetic energy dissipation rate fields are given in units of $U^2_{\rm f}$ and $U^3_{\rm f}/H$ respectively.}
	\label{fig:eps_KE}
\end{figure*}

\subsection{Temperature dependence of material parameters}
The dependence of the material parameters causes additional non-Boussinesq effects. In most applications, the temperature dependence is more relevant as the pressure dependence, see e.g. refs. \cite{Yik2020,Macek2023} for discussion. The impact of temperature-dependent thermal conductivity $k$ and dynamical viscosity $\mu$, on the dynamics of a fully compressible turbulent convection flow beyond the anelastic limit was systematically studied in ref. \cite{Panickacheril2024} by two series of three-dimensional DNS at a moderate $Ra=10^5$ and for $Pr=0.7$ in a layer of aspect ratio 4 with periodic boundary conditions in both horizontal directions. One simulation series is for a weakly stratified adiabatic background at $D=0.1$ one for a strongly stratified one at $D=0.8$.  The temperature dependence of material parameters is imposed as a power law, 
\begin{equation}
k(T)=k_0 \left(\frac{T}{T_{\rm bot}}\right)^{\beta}\quad\mbox{and}\quad \mu(T)=Pr \frac{k_0}{c_p} \left(\frac{T}{T_{\rm bot}}\right)^{\beta}
\label{eq:material}
\end{equation}
with an exponent $\beta$ that varied $0\le \beta\le 7$ for $D=0.1$ and $0\le \beta \le 0.165$ for $D=0.8$. Note that both material parameters are varied such that the Prandtl number remains constant across the layer. Equations \eqref{eq:material} result in a constant Prandtl number $Pr$ across the layer. They also cause a height-dependent superadiabaticity $\varepsilon(x_3)$. Central statistical quantities of the flow, such as the mean superadiabatic temperature, temperature and density fluctuations, or turbulent Mach numbers are compared in the form of vertical profiles averaged with respect to time and cross section plane $A=L^2$. It is found that the additional material parameter dependence causes systematic quantitative changes of all these quantities, but no qualitative ones. A growing temperature power law exponent $\beta$ also enhances the turbulent momentum transfer in the weak stratification case by 40\%; it reduces the turbulent heat transfer by up to 50\% in the strong stratification case. 

The present studies could not explore the impact of the temperature dependencies on the characteristic scale of the LLFS since the aspect ratio of our compressible convection systems was too small. Furthermore, the impact of a reduction of the Prandtl number would be desirable when keeping in mind that for solar convection $k(T)\sim T^3$ \cite{Schumacher2020}. First steps into this direction have been made in Boussinesq setups of MC with a temperature-dependent thermal diffusivity $\kappa(T)$ \cite{Shcheritsa2018,Pandey2021}. This extension did not alter the characteristic scale of the LLFS. More investigations along these lines are a part of future work.

\section{Impact of compressibility on small-scale statistics}
\label{sec6}

The dissipation rate fields take a central role in the dynamics of three-dimensional fluid turbulence, including thermally-driven turbulent convection. In the incompressible flow case, their statistical mean values determine the crossover scales from the viscosity- or diffusion-dominated to the inertia-dominated dynamics and thus set the endpoint of the near loss-free transfer of kinetic energy or scalar variance from large- to small-scale fluid flow structures \citep{Kolmogorov1941,Batchelor1959,Corrsin1951}. The local statistical analysis of dissipation rates reveals fields which vary strongly in space and time. This led to the refined similarity hypotheses by \citet{Kolmogorov1962} and \citet{Obukhov1962} in case of the velocity field. Later \citet{Stolovitzky1995} adapted this framework to the passive scalar case and the corresponding scalar dissipation rate. The strength and distribution of the dissipation rate fields have been probed by the multifractal formalism which generalizes a monofractal analysis and quantifies the uneven distribution of dissipation rate on fractal level sets \citep{Falconer2003,Grassberger1983,Hentschel1983}. Applications of this analysis for the kinetic energy dissipation rate field  \citep{Sreenivasan1991,Mukherjee2024} or the scalar dissipation rate field \citep{Schumacher2003,Schumacher2005} have been mostly done for homogeneous isotropic box turbulence. Of central importance is the kinetic energy dissipation rate field $\epsilon({\bm x},t)$ since it is closely connected to small-scale intermittency of turbulence, the anomalous scaling of velocity increment moments in the inertial cascade range $\eta_K\ll r \ll L$, and the dissipative anomaly \citep{Sreenivasan2025}. In refs. \cite{Gotoh2022,Gotoh2023} the functional form of the probability density function (PDF) of $\epsilon$ was predicted and confirmed in DNS of isotropic box turbulence. 

While the incompressible flow case has been explored in detail in the past decades, a surprisingly small number of studies was conducted on the local statistics of dissipation rates in compressible turbulence. Sarkar and co-workers \citep{Sarkar1991,Sarkar1992} decomposed mean kinetic energy dissipation rates into solenoidal and dilatational contributions, explored the asymptotic dilatational contributions in homogeneous turbulence and compared their results with DNS of isotropic compressible turbulence. These studies were extended by \citet{Huang1995} and \cite{Pirrozoli2004}, see also \citet{Pirrozoli2011}. Analyses of the dissipative anomaly in compressible isotropic turbulence followed only recently \cite{Panickacheril2021}. 

In the following, we report the statistics of the kinetic energy dissipation rate field in the bulk of a fully compressible turbulent convection flow at a Prandtl number $Pr$ of order unity. Furthermore, we investigate the statistical properties of these three components of the kinetic energy dissipation (solenoidal, dilatational and inhomogeneous) and their relative magnitudes as well as their multifractal properties. Figure~\ref{fig:eps_KE} compares components of the turbulent kinetic energy (TKE) and the kinetic energy dissipation rate of the FCC case to the corresponding quantities of the OB case \citep{Chilla2012,Samuel2024}. Note that the analysis is focused to the FCC case for which we obtained the highest turbulent Mach numbers as discussed in section 5. Following from the decomposition \eqref{Helmholtz}, we defined 
\begin{equation}
k_s({\bm x},t) = \frac{1}{2}{\bm u}^2_s \quad\mbox{and}\quad k_d({\bm x},t) = \frac{1}{2}{\bm u}^2_d\,. 
\end{equation}
The dissipation due to both the components, solenoidal and dilatational, (which will be detailed in subsection 6.1) in FCC are concentrated in the pre-shock regions, while the kinetic energy from both components is concentrated in the areas where pre-shocks tend to cluster. It is also seen that the highest amplitude structure of $\epsilon_s$ differs for OB and FCC, which suggests that the solenoidal component is also affected by the pre-shock structures.

\begin{table*}
\begin{center}
\begin{tabular}{lcccccccccc}
\hline
$\mathrm{Identifier}$  & $Ra$   &   $\varepsilon$ & $D$ & Computational grid & $Re$ & $Re_b$ & $Re_{\lambda}$ & $Re_{\lambda, b}$ & $u_{\mathrm{rms}}$ & $u_{\mathrm{rms}, b}$ \\
\hline
$\mathrm{FCC}1$   & $10^5$ & $0.45$ & $0.5$  & $256 \times 256 \times 128$ &  $79 \pm 2$ & $37 \pm 1$ & $7 \pm 0.2$ & $14 \pm 0.5$ & $0.24$ & $0.26$\\
$\mathrm{FCC}2$   & $10^6$ & $0.45$ & $0.5$ &  $512 \times 512 \times 256$ & $269 \pm 4$ & $156 \pm 3$ & $17 \pm 0.4$ & $24 \pm 0.8$ & $0.24$ & $0.27$\\
$\mathrm{OB}1$   & $10^5$ & $\to 0$ & $\to 0$ & $100 \times 100 \times 64 \times 5^3$&  $87 \pm 2$ & $58 \pm 2$ & $19 \pm 1$ & $37 \pm 2$ & $0.23$ & $0.26$\\		
$\mathrm{OB}2$  & $10^6$ & $\to 0$ & $\to 0$ & $100 \times 100 \times 64 \times 7^3$&  $285 \pm 6$ & $182 \pm 3$ & $46 \pm 1$ & $71 \pm 3$ & $0.24$ & $0.25$\\
\hline
\end{tabular}
\caption{Direct numerical simulation (DNS) in the fully compressible case (FCC) and Oberbeck-Boussinesq (OB) limit. We list Rayleigh number $Ra$, superadiabaticity $\varepsilon$, dissipation number $D$, grid resolution in mesh cells, large-scale Reynolds number $Re$, large-scale Reynolds number in the bulk $Re_b$, Taylor micro-scale Reynolds number $Re_{\lambda}$, Taylor micro-scale Reynolds number in the bulk $Re_{\lambda,b}$, root-mean-square (rms) velocity $u_{\mathrm{rms}}$, and rms velocity in the bulk $u_{\mathrm{rms}, b}$. All simulations have a Prandtl number $Pr = \nu/\kappa = 0.7$ and aspect ratio $\Gamma = 4$. The corresponding Taylor microscale Reynolds numbers are calculated as $Re_{\lambda}=\sqrt{5 \langle \rho\rangle_{V,t}/(3\langle\epsilon\rangle_{V,t} \mu)}\, u^2_{\rm rms}$; for $Re_{\lambda,b}$ the volume $V$ is substituted by the bulk volume $V_b=L^2\times [0.2,0.8] H$ in OB1, OB2, FCC2 and $V_b=L^2\times [0.2,0.6] H$ in FCC1. Velocities are given in units of $U_{\rm f}=\sqrt{\varepsilon g H}$. Data of runs FCC1 and FCC2 are obtained by a CFDM, OB1 and OB2 by a SEM.}
\label{tab:tab1}
\end{center}
\end{table*}
\begin{table*}
\begin{center}
\begin{tabular}{lcccccccccc}
\hline
$\mathrm{Identifier}$  & $M_t$ & $M_{t,b}$ & $M_{t}^{\mathrm{max}}$ & $M_{t,b}^{\mathrm{max}}$ & $\delta$ & $\delta_b$ & $\delta^{\mathrm{max}}$ & $\delta_b^{\mathrm{max}}$ \\
\hline
$\mathrm{FCC}1$   & $0.23$ & $0.22$ & $ 1.1 $ & $ 1.1 $ & $0.35$  & $0.28$ & $0.38$ & $0.31$ & \\
$\mathrm{FCC}2$    & $0.24$ & $0.24$ & $1.3$  & $1.3$ & $0.31$ & $0.27$ &  $0.33$ & $0.28$ & \\
\hline
\end{tabular}
\caption{Further parameters for the fully compressible convection runs FCC1 and FCC2. We list mean turbulent Mach number $M_t$, mean turbulent Mach number in the bulk $M_{t,b}$, maximum turbulent Mach number $M_t^{\mathrm{max}}$, maximum turbulent Mach number in the bulk $M_{t,b}^{\mathrm{max}}$, mean dilatational parameter $\delta$, mean dilatational parameter in the bulk $\delta_b$, maximum dilatational parameter $\delta^{\mathrm{max}}$, maximum dilatational parameter in the bulk $\delta_b^{\mathrm{max}}$. Note that the maximum turbulent Mach number occurs in the bulk.}
\label{tab:tab2}
\end{center}
\end{table*}
\begin{table*}
\begin{center}
\begin{tabular}{lcccccccccc}
\hline
$\mathrm{Identifier}$  & $\langle\epsilon\rangle_{V,t}$ & $\langle\epsilon\rangle_{V_b,t}$ & $\langle \epsilon_s \rangle_{V,t}$ & $\langle \epsilon_{s} \rangle_{V_b,t}$ &  $\langle \epsilon_{d} \rangle_{V,t}$ & $\langle \epsilon_{d} \rangle_{V_b,t}$ &  $\langle \epsilon_{I} \rangle_{V,t}$ & $\langle \epsilon_{I} \rangle_{V_b,t}$ \\
\hline
$\mathrm{FCC}1$   & $3.1 \times 10^{-2}$  & $1.4 \times 10^{-2}$ & $2.8 \times 10^{-2}$ & $1.7 \times 10^{-2}$ & $3.8 \times 10^{-4}$ & $5.1 \times 10^{-4}$ & $3.0 \times 10^{-3}$ & $-3.5 \times 10^{-3}$ \\
$\mathrm{FCC}2$   & $2.1 \times 10^{-2}$ &  $1.4 \times 10^{-2}$ & $2.0 \times 10^{-2}$ & $1.5 \times 10^{-2}$ & $1.9 \times 10^{-4}$ & $2.3 \times 10^{-4}$ & $1.3 \times 10^{-3}$ & $-1.1\times 10^{-3}$\\
$\mathrm{OB}1$   & $4.6 \times 10^{-3}$ &  $2.1 \times 10^{-3}$ & $4.3 \times 10^{-3}$ & $2.6 \times 10^{-3}$ & -- & -- & -- & --\\
$\mathrm{OB}2$    & $3.0 \times 10^{-3}$ &  $1.6 \times 10^{-3}$ & $2.9 \times 10^{-3}$ & $1.7 \times 10^{-3}$ & -- & -- & -- & --\\
\hline
\end{tabular}
\caption{Mean values of the total dissipation rate and its three components, cf. eq. \eqref{eq:disstotal}, in the full domain $V$ and the bulk $V_b$ for all four runs. Dissipation rates are given in units of $U_{\rm f}^3/H$.}
\label{tab:tab3}
\end{center}
\end{table*}

\subsection{Decomposition of kinetic energy dissipation rate} 
We decompose the kinetic energy dissipation rate field of kinetic energy into three parts, the solenoidal and dilatational components and an additional inhomogeneous component~\citep{Huang1995} which is absent in isotropic box turbulence. Take the dot product of the momentum equation (\ref{eq:mom}) with $u_i$ and get a balance equation of the local kinetic energy which is given by
\begin{equation}
\frac{\partial}{\partial t} \left( \rho  \frac{u_i^2}{2} \right) + \frac{
\partial}{\partial x_j} \left( \rho u_j \frac{u_i^2}{2} + u_j p - u_i \sigma_{ij} \right) = p \frac{\partial u_i}{\partial x_i}  - \sigma_{ij}\frac{\partial 
u_{i}}{\partial x_j}  - \rho g u_i \delta_{i,3} \,.
\label{eq:ke_balance}
\end{equation}
The terms on the right-hand side of (\ref{eq:ke_balance}) represent the pressure dilatation, energy dissipation rate, and energy injection by volume forcing. We denote the kinetic energy dissipation rate field (per unit mass) by $\epsilon$ and obtain
\begin{align}
\epsilon({\bm x},t)  &= \frac{\sigma_{ij}}{\rho}\frac{\partial u_{i}}{\partial x_j} =  2\nu S_{ij}\frac{\partial u_{i}}{\partial x_j}   - \frac{2\nu}{3} \left( \frac{
\partial u_k}{\partial x_k} \right)^2 \,,
\label{eq:viscous_dissi}
\end{align}
where $\nu = \mu/\rho$ is the kinematic viscosity. The velocity gradient tensor can be decomposed into $\partial u_{i}/\partial x_j = S_{ij} +  \Omega_{ij}$, where $\Omega_{ij} = (\partial u_{i}/\partial x_j-\partial u_{j}/\partial x_i)/2$ is the anti-symmetric vorticity tensor and $S_{ij} \Omega_{ij} = 0$. Thus, the first term on the right-hand side of (\ref{eq:viscous_dissi}) equals $2\nu S_{ij} S_{ij}$, which can be expanded as follows,
\begin{align}
2\nu S_{ij}S_{ij} = 2\nu \Omega_{ij}\Omega_{ij} + 2\nu \left[ \frac{\partial u_i}{\partial x_j}  \frac{\partial u_j}{\partial x_i} \right] \,.
\label{eq:sijsij}
\end{align}
Using the product rule, we express the second term on the right hand side of (\ref{eq:sijsij}) as
\begin{align} \nonumber
2\nu \left[  \frac{\partial u_i}{\partial x_j}  \frac{\partial u_j}{\partial x_i} \right] &= 2\nu \left[ \frac{\partial}{\partial x_j} \left\{ u_i \frac{\partial u_j}{\partial x_i} 
	\right\}  - u_i\frac{\partial }{\partial x_j} \left( \frac{\partial u_j}{\partial x_i} \right) \right] \\ \nonumber
&= 2\nu  \frac{\partial }{\partial x_j}  \left\{ \frac{\partial }{\partial x_i} (u_iu_j) - u_j \frac{\partial u_i }{\partial x_i}\right\} \nonumber \\ &\hspace{30pt} - 2\nu \left\{\frac{\partial }{\partial x_i} \left( u_i \frac{\partial u_j}{\partial x_j} \right) - \left( \frac{\partial u_j}{\partial x_j} \right)^2 \right\} \nonumber \\
&= 2\nu \frac{\partial^2}{\partial x_i\partial x_j} \left( u_i u_j\right) - 4\nu \frac{\partial}{\partial x_i} \left( u_i \frac{\partial u_k}{\partial x_k}\right) + 2\nu \left( \frac{\partial u_k}{\partial x_k} \right)^2 \,.
\label{eq:der_product}
\end{align}
Combining eqns. \eqref{eq:viscous_dissi}, \eqref{eq:sijsij}, and \eqref{eq:der_product} yields the decomposition, 
\begin{equation}
\epsilon =  \epsilon_s + \epsilon_d + \epsilon_I\,,
\label{eq:disstotal}
\end{equation}
with the solenoidal (s), dilatational (d), and inhomogeneous (I) components, which follow to
\begin{align}
\epsilon_s({\bm x},t)  &=  2 \nu \Omega_{ij}\Omega_{ij}\,,\nonumber\\ \epsilon_d({\bm x},t)  &= \frac{4\nu}{3}\left( \frac{
\partial u_k}{\partial x_k} \right)^2\,, \nonumber \\
\epsilon_I({\bm x},t)  &= 2\nu \left[ \frac{\partial^2}{\partial x_i\partial x_j} \left( u_i u_j\right)  - 2 \frac{\partial}{\partial x_i} \left(u_i \frac{\partial u_k}{\partial x_k}\right) \right] \,.
\end{align}
Using the definition of vorticity vector field $\omega_i = -\varepsilon_{ijk} \Omega_{jk}$, where $\varepsilon_{ijk}$ is the Levi-Civita symbol, it can be shown that the local enstrophy $\omega^2  = \omega_i \omega_i= 2 \Omega_{jk} \Omega_{jk}$. Thus, $\epsilon_s =  2 \nu \Omega_{ij}\Omega_{ij} = \nu \omega^2$, and thence, for constant $\nu$, $\epsilon_s$ is directly proportional to the local enstrophy, $\omega^2$. The statistical properties of the three components in the bulk of the convection layer will be studied now. Furthermore, a comparison to the OB cases at the same Rayleigh numbers is provided. Table \ref{tab:tab1} summarizes the DNS. Runs FCC1 and FCC2 are conducted at $Ra=10^5$ and $Ra=10^6$ and compared to OB1 and OB2, respectively. Essential simulation parameters are provided in the table. Table \ref{tab:tab2} lists in addition the averaged turbulent Mach numbers $M_t$, cf. eq. \eqref{eq:Mt}, and the dilatational parameter $\delta$ which is given by
\begin{equation}
\delta=\frac{u_{d,{\rm rms}}}{u_{s,{\rm rms}}}\,.
\end{equation}
In ref. \cite{Donzis2020}, it was shown, that $M_t$ and $\delta$ have to be considered together with $Re_{\lambda}$ when investigating universal scaling properties of compressible small-scale turbulence.

\begin{figure*}[ht!]
\centering
\includegraphics[width=0.72\linewidth]{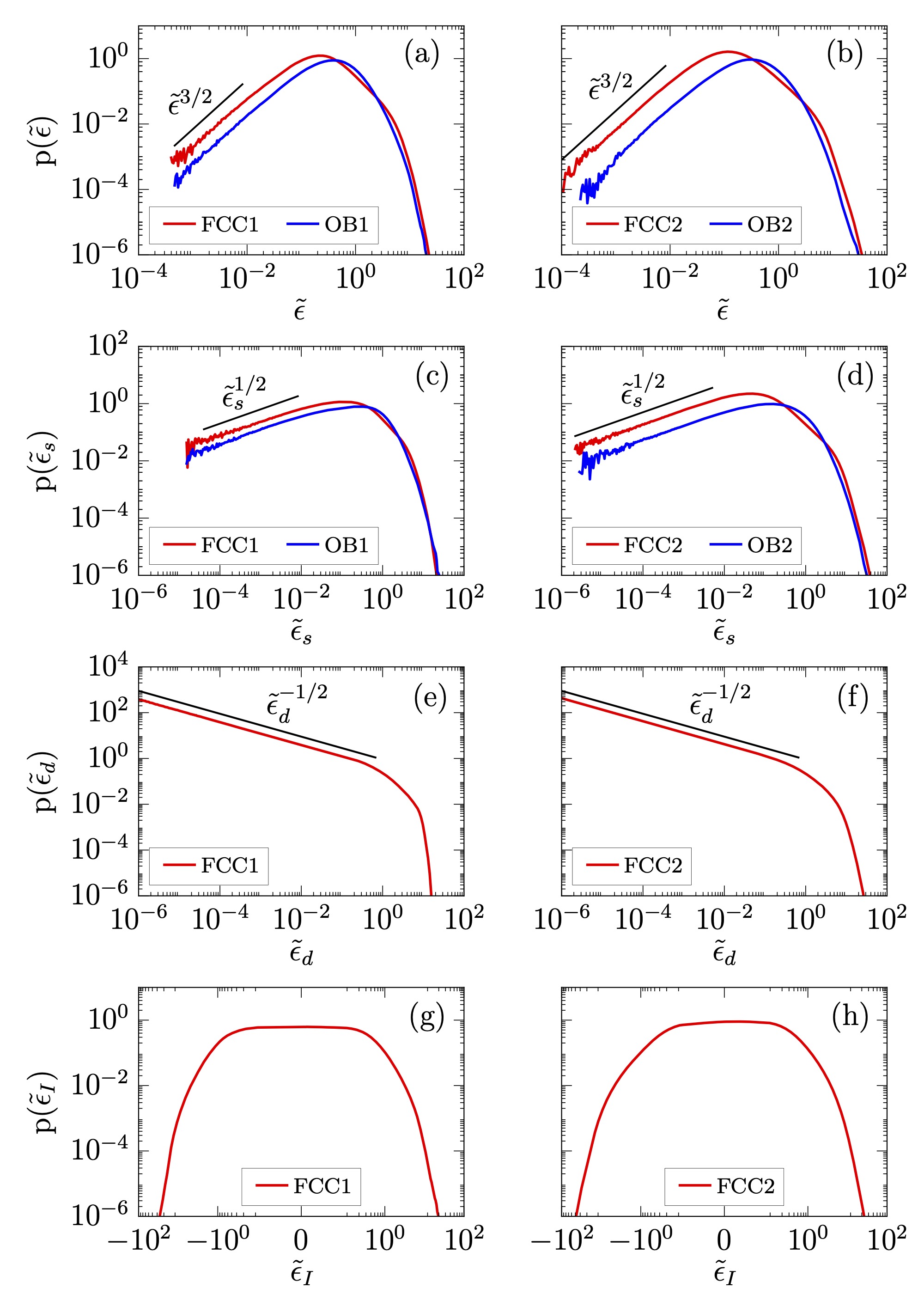}
\caption{Probability density functions (PDFs) of different components of the kinetic energy dissipation rate in a doubly logarithmic plot. Data are shown for OB and FCC in the bulk of the convection layer. The left column is for $Ra=10^5$, the right one for $Ra=10^6$. (a, b) PDF of the total field, $\mathrm{P}(\tilde{\epsilon})$. (c, d) PDF of $\mathrm{P}(\tilde{\epsilon}_s)$. (e, d) PDF of $\mathrm{P}(\tilde{\epsilon}_d)$. (g, h) PDF of $\mathrm{P}(\tilde{\epsilon}_I)$. In all cases, the normalized fields are shown, $\tilde{\epsilon}_k({\bm x},t) = \epsilon_k({\bm x},t) / \langle \epsilon_k \rangle_{V_b,t} $ for $k=\{s,d,I\}$ and the total dissipation.}
\label{fig:pdfs}
\end{figure*}
\begin{figure*}[ht!]
\centering
\includegraphics[width=0.85\linewidth]{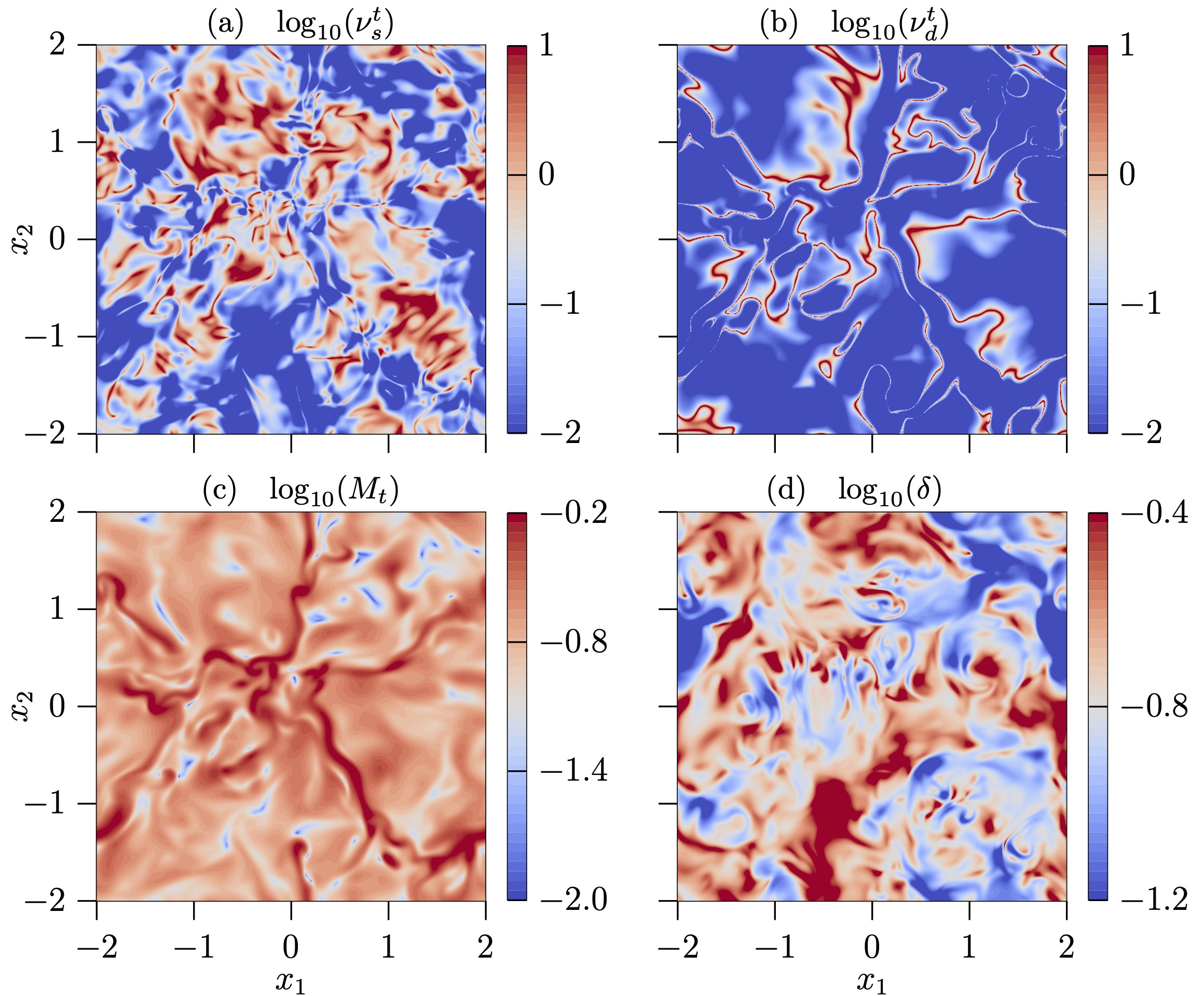}
\caption{Snapshot contour plots of different quantities taken at half height $x_3=H/2$ of the convection layer height for FCC2 at $Ra=10^6$ and $Pr=0.7$. (a) Solenoidal turbulent viscosity field $\nu^t_s$. (b) Dilatational turbulent viscosity field $\nu^t_d$. (c) Local turbulent Mach number $M_t$. (d) Local dilatational parameter $\delta$.  All fields and corresponding color bars are given in logarithmic levels. Turbulent viscosity fields are given in units of $U_{\rm f}H$.}
\label{fig:nu_Mt_delta}
\end{figure*}
\begin{figure}[ht!]
\centering
\includegraphics[width=0.85\linewidth]{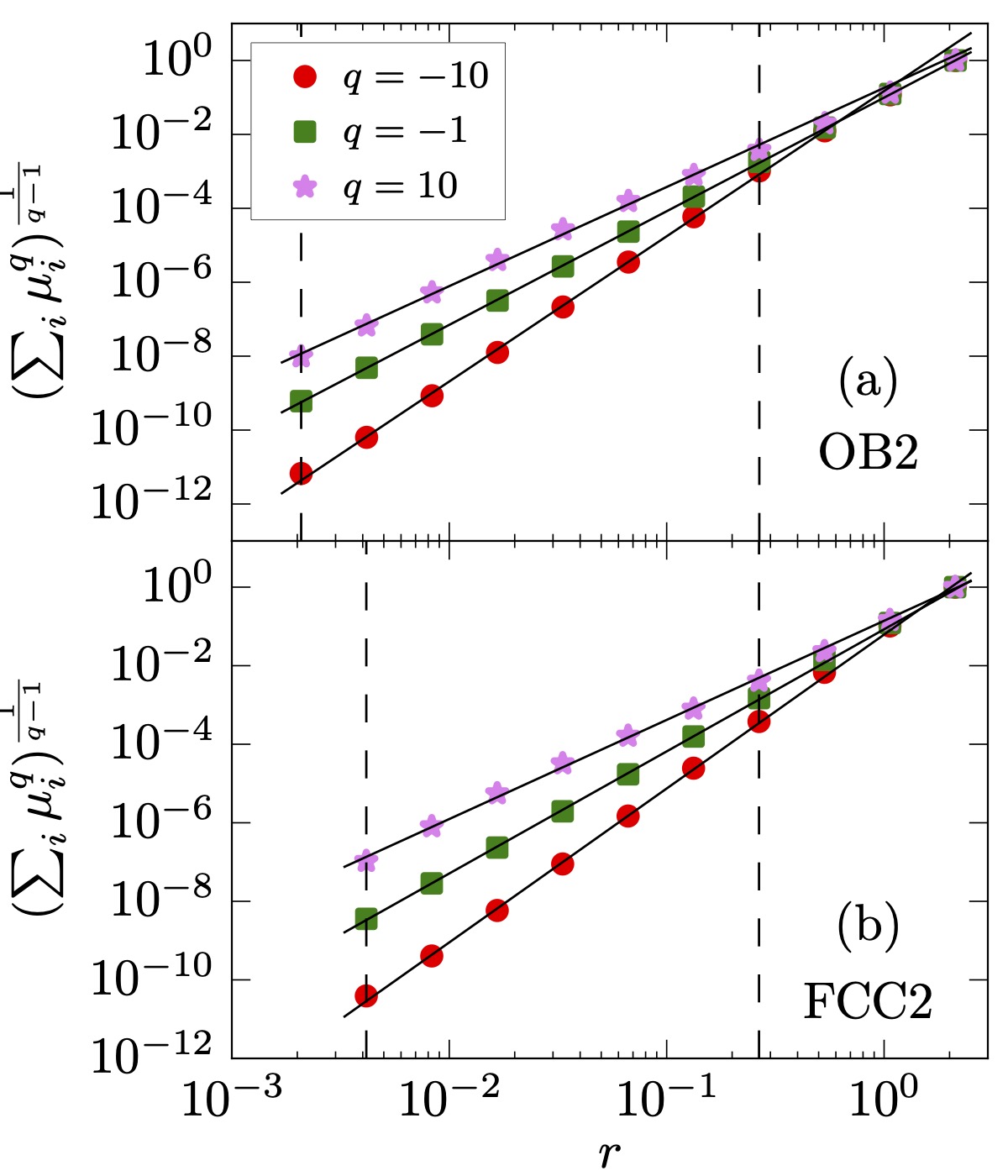}
\caption{Double logarithmic plots of the sum of the moments of the coarse-grained energy dissipation field, $(\sum_i \mu_i^q)^{1/{q-1}}$ versus $r$. Data are shown for powers of $q = \{ -10, -1, 10\}$ to illustrate the scaling range for the evaluation of the corresponding generalized dimensions, $D(q)$ by a power law fit. (a) Case OB2 and (b) case FCC2. The black solid lines show the best fits. Vertical dashed lines represent the fit range of $r$.} 
\label{fig:exponents}
\end{figure}
\begin{figure*}[ht!]
	\centering
	\includegraphics[width=0.85\linewidth]{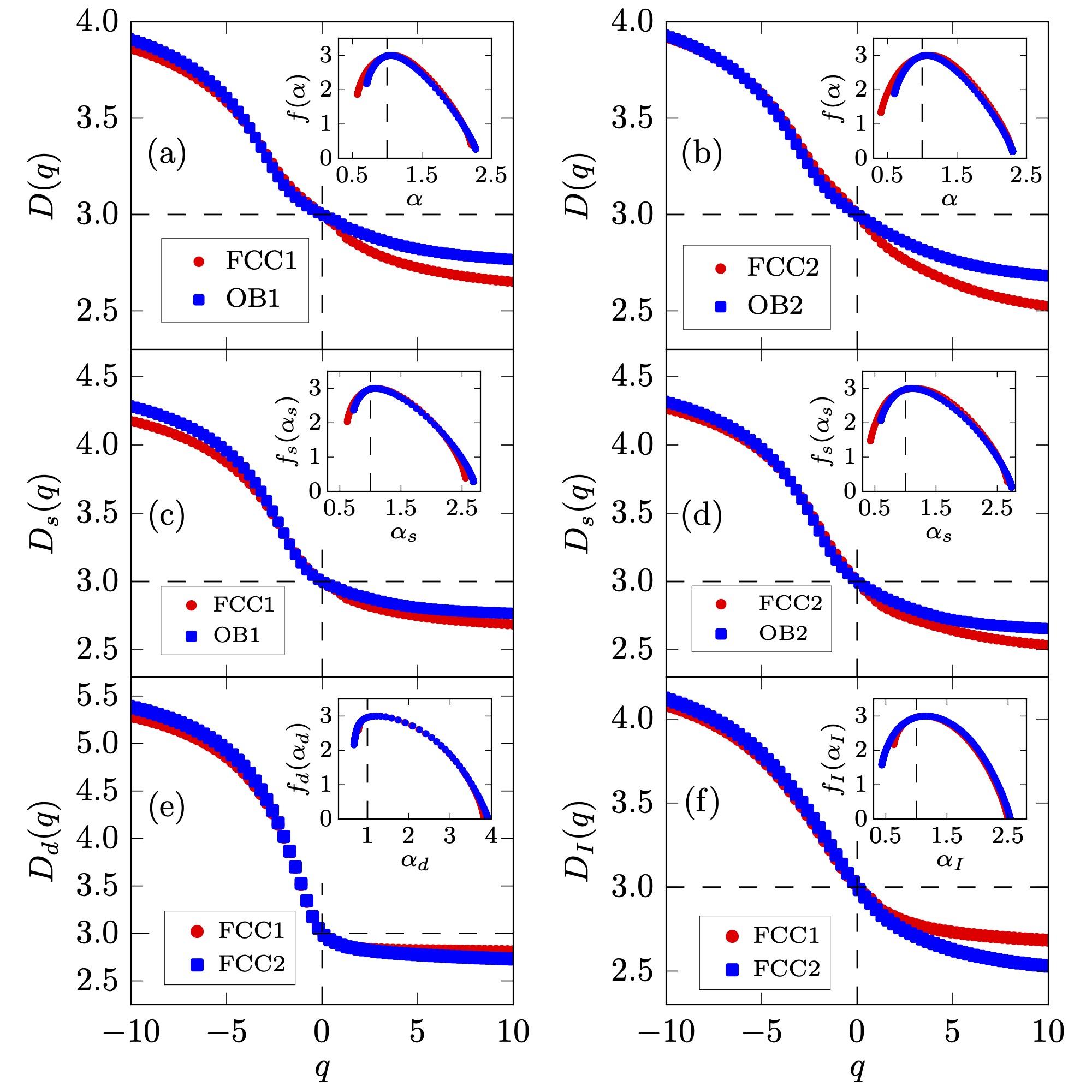}
	\caption{Generalized dimensions  (main panels) and related singularity spectra (insets) are shown for different components of the dissipation rate field. (a, b) $D(q)$ and $f(\alpha)$ of the total dissipation rate field $\epsilon$. (c, d) $D_s(q)$ and $f_s(\alpha_s)$ of the solenoidal dissipation rate field $\epsilon_s$. (e) $D_d(q)$ and $f_d(\alpha_d)$ of the dilatational dissipation rate field $\epsilon_d$. (f) $D_I(q)$ and $f_I(\alpha_I)$ of the absolute inhomogeneous dissipation rate field $|\epsilon_I|$. The vertical dashed lines in the insets correspond to $\alpha = 1$.}
	\label{fig:Dqs}
\end{figure*}

\subsection{Probability density functions}
Figure \ref{fig:pdfs} shows probability density functions (PDFs) of all kinetic energy dissipation rate components together with the total one, cf. eq. \eqref{eq:disstotal} in doubly logarithmic plots for FCC1 and FCC2. The wide tails of the PDFs of the dissipation field highlight the occurrence of both, low- and high-amplitude dissipation events, which are characteristic for non-Gaussian statistics observed in turbulent flows. To this end,  we examine the PDFs of the normalized fields $\tilde{\epsilon}_k({\bm x},t) = \epsilon_k({\bm x},t) / \langle \epsilon_k \rangle_{V_b,t} $ for $k=\{s,d,I\}$ and the total field. The bulk of convection layer is $V_b=L^2\times [0.2,0.8]H$ for OB1, OB2, FCC2 and $V_b=L^2\times [0.2,0.6]H$ for run FCC1. Data are compared to the corresponding OB case at the same Rayleigh number if possible. The following observations can be made:
\begin{enumerate}
\item As evident from Figs.~\ref{fig:pdfs}(a-d), occurrences of $\epsilon < \langle \epsilon \rangle_{V_b,t}$  and $\epsilon_s < \langle \epsilon_s \rangle_{V_b,t}$ are more frequent in FCC then in OB convection, and this difference between both flows in the left tail increases with increasing $Ra$. 
\item The PDFs $\mathrm{P}(\tilde{\epsilon})$ and $\mathrm{P}(\tilde{\epsilon}_s)$ in FCC exhibit somewhat fatter right tails which clearly visible in Fig.~\ref{fig:pdfs}(b). This suggests that the compressible case shows an enhanced intermittency, despite FCC has a slightly lower Reynolds number compared to OB. Further below in subsection 6.3, we will demonstrate this enhanced intermittency in FCC by the multifractal analysis. 
\item The left tails of $\mathrm{P}(\tilde{\epsilon})$ and $\mathrm{P}(\tilde{\epsilon}_s)$ exhibit a scaling law of $\tilde{\epsilon}^{3/2}$ and $\tilde{\epsilon}_s^{1/2}$ respectively, mirroring the same power laws observed in incompressible homogeneous isotropic turbulence (HIT)~\citep{Gotoh2022,Gotoh2023}. We detect this scaling for OB and FCC. 
\item Similar to the trend observed in incompressible HIT, the peaks of $\mathrm{P}(\tilde{\epsilon})$ and $\mathrm{P}(\tilde{\epsilon}_s)$ shift toward lower $\tilde{\epsilon}$ and $\tilde{\epsilon}_s$, respectively, as the Rayleigh number increases. 
\item Compressibility leaves a noticeable imprint on the shape of the transient region between the two tails. This region is qualitatively similar for OB to that of HIT. 
\item The distribution $\mathrm{P}(\tilde{\epsilon}_d)$ in Figs.~\ref{fig:pdfs}(e,f) decreases monotonically with increasing $\tilde{\epsilon}_d$, adhering to a $\tilde{\epsilon}_d^{-1/2}$ power law when $\tilde{\epsilon}_d < 1$, whereas decreasing very rapidly for $\tilde{\epsilon}_d > 1$. 
\item The inhomogeneous component is not positive definite and the resulting PDF exhibits a longer left tail, compared to the right one, as shown in Figs.~\ref{fig:pdfs}(g,h). 
\end{enumerate}
Table \ref{tab:tab3} displays in addition the full volume and bulk volume means of all dissipation components.

Figure \ref{fig:nu_Mt_delta} shows horizontal midplane cross sections of a snapshot of the solenoidal and dilatational turbulent viscosity fields together with the turbulent Mach number and the dilatational parameter. All fields are obtained by a pointwise and instantaneous evaluation, shown for simulation run FCC2. In detail, we adapted the definition of the $M_t$ and $\delta$ to 
\begin{equation}
M_t({\bm x},t_0)=\sqrt{\frac{{\bm u}^2}{\gamma R T}}\quad \mbox{and}\quad \delta({\bm x},t_0)=\sqrt{\frac{{\bm u}^2_d}{{\bm u}^2_s}}\,.
\end{equation}
The turbulent viscosity is a widely used measure of the strength of turbulent mixing in a flow \cite{Pope2000,Pandey2022a}. The standard definition is given by $\nu_t=C_{\nu}\langle k\rangle /(2\langle \epsilon\rangle)$ where $C_{\nu}\approx 0.09$. Here, $\langle\cdot\rangle$ denotes an appropriately taken average. We adapted this standard definition to a pointwise one, which is given by
\begin{equation}
\nu_s^t({\bm x},t_0)=\frac{k_s^2}{\epsilon_s}\quad \mbox{and}\quad \nu_d^t({\bm x},t_0)=\frac{k_d^2}{\epsilon_d}\,,
\end{equation}
and skip the unknown prefactors $C_{\nu}/2$ for this qualitative comparison. Both contributions to a turbulent viscosity vary over the same range of amplitudes in the present example. Furthermore, the boundaries of regions of enhanced turbulent Mach number $M_t$ are local maxima of $\nu_d^t$. When comparing the contours of $\nu_s^t$ and $\nu_d^t$, we confirm an enhanced turbulent mixing by the solenoidal component $\nu_s^t$ in regions outside the pre-shocks. These regions of enhanced turbulent viscosity partly overlap with local maxima of the dilatational parameter. It remains to be seen how this behavior changes for higher Rayleigh numbers and stronger degrees of stratification of the adiabatic equilibrium ($D\to 1$).

\subsection{Multifractal analysis of the kinetic energy dissipation rate}
In the limit of very large Reynolds number, the incompressible three-dimensional Navier-Stokes equations are invariant under the scaling transformations \cite{Frisch1994}
\begin{align}
{\bm r} &\rightarrow \lambda {\bm r}\,, \nonumber\\ 
{\bm u} &\rightarrow \lambda^{\alpha / 3}{\bm u}\,,\nonumber\\ 
t &\rightarrow \lambda^{1-\alpha/3} t\,,
\label{eq:scaling}
\end{align}
for scaling factor $\lambda > 0$; $\alpha$ is an arbitrary scaling exponent. This scaling symmetry is broken when the flow is in the vicinity of walls where viscous effects become important. By dimensional arguments, transformations \eqref{eq:scaling} lead to a scaling of the kinetic energy dissipation rate, that is given by 
\begin{equation}
\epsilon_r({\bm x}, t)=\frac{1}{B(r)}\int_{B(r)}\epsilon({\bm x}+{\bm r},t)d^3 r \sim r^{\alpha-1}\,.
\label{eq:scaling1}
\end{equation}
The field is averaged over a small subvolume (or ball) with size (or radius) $r$. Local isotropy is assumed in addition in \eqref{eq:scaling1} and $|{\bm r}|=r$, the coarse-graining scale. The classical theory of Kolmogorov \cite{Kolmogorov1941,Frisch1994} (also denoted as K41) assumes that the averaged kinetic energy dissipation field is independent of the size of the averaging domain $B(r)$, which corresponds to $\alpha = 1$ in eq. \eqref{eq:scaling1}. However, an essential feature of dissipation fields are the highly intermittent behaviour, both in space and time. Several intermittency models~\citep{Kolmogorov1962, Novikov1964, Mandelbrot1974, Frisch1978}, see also ref. \cite{Sreenivasan1997,Sreenivasan2025} for reviews, have been proposed to describe the resulting scaling properties of the dissipation field and connect them to the scaling of increment moments of the velocity field ${\bm u}$. 

The multifractal framework considers the strongly fluctuating  energy dissipation field as a highly unevenly in space distributed measure. Each subset follows a particular local scaling $r^\alpha$ and is spatially supported on a monofractal subset of the three-dimensional space with a dimension $f(\alpha)$ . All these differently scaling subsets are interwoven in a turbulent flow and the set of corresponding exponents $\alpha$ forms the \textit{singularity spectrum}. The strongest singularities are connected to $\alpha\to 0$, i.e. for exponents smaller than $\alpha=1$ following from K41. The multifractal model of energy dissipation has been also verified in laboratory experiments, e.g. \cite{Meneveau1987, Sreenivasan1991}. This framework is now applied to the compressible convection in the bulk of the convection layer away from the top and bottom walls to investigate the spatially intermittent distribution of the components of the kinetic energy dissipation rate. \citet{Halsey1986} determined the singularity strength $\alpha$ and the associated fractal dimensions $f(\alpha)$ by connecting them to a hierarchy of \textit{generalized dimensions} $D(q)$, which were introduced in \cite{Hentschel1983}. They are given by
\begin{equation}
D(q) =  \lim_{r \rightarrow 0} \frac{1}{q-1} \frac{\log \sum_i \mu_i^q(r)}{\log r},
\label{eq:Dq}
\end{equation}
where power $q \in (-\infty, +\infty)$ and the measure $\mu_i(r)$ correspond to subvolume $B_i(r)$ and are calculated as follows, 
\begin{equation}
\mu_i(r) = \frac{\mathcal{E}_{r}}{\mathcal{E}} \quad \mbox{with} \quad \bigcup_{i=1}^{N_r} B_{i}(r)=V_b\,,
\end{equation}
with
\begin{equation}
\mathcal{E}_{r} = \epsilon_r r^d \sim r^{\alpha -1 +d}\quad\mbox{and}\quad
\mathcal{E} =\langle \epsilon\rangle_{V_b} V_b\,,
\label{eq:Dq1}
\end{equation}
with $d=3$. The slope of the doubly logarithmic plots of the sum $(\sum_i \mu_i^q(r))^{1/(q-1)}$ versus $r$ determines the generalized dimension $D(q)$ for each $q$, as shown in Fig.~\ref{fig:exponents} for the runs OB2 and FCC2. In ref. \cite{Halsey1986}, it was shown that $\alpha$ and $f(\alpha)$ are connected to $q$ and $D(q)$ by a Legendre transformation, see \citet{Meneveau1987} for a detailed derivation. Exponent $\alpha$ and corresponding fractal dimension $f(\alpha)$ are given by 
\begin{align}
\alpha &= \frac{d}{dq}[(q-1)(D(q)-d+1)]\,,  \\
f(\alpha) &= \alpha q - (q-1)(D(q) -d +1) +d -1\,.
\label{eq:f_alpha}
\end{align}
Likewise, employing (\ref{eq:Dq}) and (\ref{eq:f_alpha}) for the components $\epsilon_s$, $\epsilon_d$ and $\epsilon_I$ of $\epsilon$, we compute the generalized dimensions  $D_\beta(q)$ and the singularity spectra $f_\beta(\alpha_\beta)$, where $\beta = \{s, d, I\}$ or the total dissipation rate. Note that by applying (\ref{eq:f_alpha}) the non-intermittent assumption $D(q) = d$ from K41 yields $\alpha = 1$ and $f(\alpha) = d$, which is here $d=3$. 

Figure~\ref{fig:Dqs} demonstrates that the dissipation rate fields possess a spectrum of  generalized dimensions (main panels) and fractal dimensions (insets) which significantly differ from the dimension of the embedding physical space, $d = 3$. This analysis clearly underlines the intermittent nature of all components of the kinetic energy dissipation field, even though the Rayleigh numbers are moderate only. For powers $q>0$, Figs.~\ref{fig:Dqs} (a,b) and \ref{fig:Dqs} (c,d) reveal that in case of FCC dimensions $D(q)$ and $D_s(q)$ are smaller compared to OB convection case. This leads to smaller values of the singularities  $\alpha$ and $\alpha_s$, as well as associated fractal dimensions $f(\alpha)$ and $f(\alpha_s)$ for FCC. Consequently, a higher degree of intermittency (or of ``roughness`` of the spatial distribution) in the compressible convection case is observed, despite the slightly smaller Reynolds number, which was also highlighted earlier in our discussion of the PDFs in subsection 6.2. The $D(q)$ curves for $q<0$ are dominated by the minima of dissipation fields~\citep{Schumacher2005}. In this region, $D_d(q)$ is largest thus highlighting the pronounced regions of comparatively small energy dissipation between the high-dissipation-amplitude pre-shocks, cf. panel (e) of the figure and panel (c) of Fig. \ref{fig:eps_KE}. The present analysis should be considered as a first step only which requires a continuation to higher Rayleigh numbers in order to quantify the scaling of the dissipation rate statistics with $Ra$.

\section{Conclusions and outlook}
\label{sec7}
The present work intends to review recent studies on mesoscale convection and to extend these investigations with new results. The analysis is focused to a paradigm of convection processes in nature, a plane layer configuration in which turbulent velocity and temperature fields are coupled and fluid motion is driven by buoyancy forces. We did not take couplings to further physical processes into account, such as weak or strong rotation (except for the discussion of the supergranule formation), magnetic fields, radiative transfer, or changes of the phases of the working fluid or its chemical composition. In natural flows, often some of these processes are connected to thermal convection. Our discussion is based on three-dimensional direct numerical simulations without subgrid scale modeling. Consequently they are limited in magnitude of Rayleigh and Prandtl numbers, in particular when the plane layer configuration with horizontal extensions larger than the height, $\Gamma >1$, is considered.        

Our study discusses the Oberbeck-Boussinesq as well as the non-Boussinesq regimes of thermal convection. To this end and as shown in Fig. \ref{fig:Fig2}, we investigated 4 specific aspects of MC independently of each other. These were (1) the low-Prandtl-number regime and (2) the role of boundary conditions on the formation of large-scale patterns, the LLFSs. Both aspects were studied in the OB limit of MC. We showed that thermal boundary conditions determine the structure size which can be traced back to the primary linear instability at the onset of convection. Large-scale patterns of MC are found for all Prandtl numbers that were accessible in our simulation studies, very low and very high $Pr$ \cite{Vieweg2023}. These are the TSSs for Dirichlet boundary conditions of temperature and granule and SG cells for Neumann boundary conditions. The characteristic scale of these LLFSs depends weakly on $Pr$ and $Ra$ in the Dirichlet case \cite{Stevens2018,Pandey2018,Krug2020}; in summary, the characteristic pattern scale grows slightly at fixed $Ra$ when $Pr$ is increased. Independent of the Prandtl number is the ratio of local coherent shear-dominated and incoherent shear-free velocity field regions near walls in case of thermal Dirichlet conditions \cite{Samuel2024}. The fraction of near-wall coherent flow $A_{\rm coh}$ is always about 40\%. The LLFSs, which typically take the form of circulation rolls, act as barriers to material transport in convection \cite{Schneide2022}; they are also regions of reduced heat transfer when compared to the flow regions between the circulation roles that strongly mix fluid.  

Aspects (3) and (4) from Fig. \ref{fig:Fig2} are related to non-Boussinesq thermal convection beyond the OB and anelastic limits. Here, we discussed the fully compressible regime of thermal convection, identified several convection regimes that differ strongly in terms of superadiabaticity, strength of stratification of the adiabatic equilibrium state $\{\bar{T}(x_3),\bar{p}(x_3),\bar{\rho}(x_3)\}$, and the resulting characteristic flow structures. The strongest asymmetry between the top and bottom boundary layer follows for the strongly stratified convection regime with $D\to 1-\varepsilon$. Here, the top boundary layer is strongly stabilized, interspersed by zones of detachment of localized coherent plumes that can sink deeply into the bulk and get further focused by compressibility. This differs fundamentally from the boundary layer dynamics in the OB regime. The downwellings are contrasted by low-amplitude broader distributed upwellings. Such a highly asymmetric top-down plume dynamics is suggested to be at work at the surface of the Sun \cite{Brandenburg2017,Schumacher2020}. Furthermore, we introduced a temperature dependence of the thermal conductivity $k(T)$ and the dynamic viscosity $\mu(T)$ \cite{Panickacheril2024}. The temperature dependence was modeled as a power law and caused quantitative changes in global heat and momentum transfer depending on the magnitude of $D$. Characteristic features of compressible turbulence, such as pre-shocks appear most prominently for a similar magnitude of superadiabaticity and dissipation number, $\varepsilon\approx D$. Note that for purely buoyancy-driven turbulence, it can be shown that the free-fall Mach number $M_f=U_f/c_s\le 1$ \cite{Panickacheril2023} and that the compressible flow remains subsonic. 

The presented decomposition of the compressible velocity field into solenoidal and dilatational components reveals a highly intermittent kinetic energy dissipation rate which results from different contributions in pre-shock-dominated regions and the spatial complements. Small-scale mixing is enhanced outside pre-shock regions, which was quantified here by the determination of an eddy or turbulent viscosity. Maxima of the dissipation field result from pre-shock regions of the dilatational component and shear layers of the solenoidal one in between. It can be expected that the pre-shock contributions become more prominent when the dilatational parameter $\delta\gtrsim 1$. This would require an additional dilatational volume forcing by additional mechanisms (and cannot be obtained by a pure buoyancy forcing of the flow).  

It is clear that our presented work should be considered as a starting point only. Future studies have to bring these 4 aspects closer together (which we did in parts already). This includes for example fully compressible convection at low Prandtl number and in extended domains to come closer to the solar granule configuration of Fig. \ref{fig:Fig1} or other stellar convection configurations. Future studies also require to further extend the range of accessible Rayleigh numbers in plane layer convection beyond the present values of $Ra \le 10^{11}$ of ref. \cite{Samuel2024} which we obtained for an aspect ratio $\Gamma=4$. The exploration of convection with DNS along these suggested lines will help to improve the parametrizations of MC in global simulation models, which typically do not include intermittent non-Gaussian statistics. Such parametrizations can be data-driven applying classical or hybrid quantum classical reservoir computing \cite{Pandey2020,Pfeffer2022,Pfeffer2023,Kobayashi2024} or generative models \cite{Heyder2024}.  This outlook brings us back to the original motivation in Fig. \ref{fig:Fig1} at the beginning of this article. Corresponding investigations are currently underway and will be reported elsewhere.     

\vspace{0.5cm}
\noindent \textbf{Acknowledgements}\\
The work of S.A., J.P.J and P.P.V. was supported by the Deutsche Forschungsgemeinschaft. J.P.J. also received support by the Alexander von Humboldt Foundation. The work R.J.S. is funded by the European Union (ERC, MesoComp, 101052786). Views and opinions expressed are however those of the author(s) only and do not necessarily reflect those of the European Union or the European Research Council. Neither the European Union nor the granting authority can be held responsible for them. Supercomputer time has been provided at the University Computer Center (UniRZ) of the TU Ilmenau. The authors also gratefully acknowledge the Gauss Center for Supercomputing e.V. (https://www.gauss-centre.eu) for funding this project by providing computing time on the GCS Supercomputers SuperMUC-NG at the Leibniz Supercomputing Center (https://www.lrz.de) and JUWELS at the J\"ulich Supercomputing Center (https://www.fz-juelich.de/en/ias/jsc). We thank Mathis Bode, Diego A. Donzis, Matti Ettel, Toshiyuki Gotoh, Janet D. Scheel, Katepalli R. Sreenivasan, and Mahendra Verma for their support and helpful discussions. J.S. wishes to thank the organizers of the 13th Turbulence and Shear Flow Phenomena Conference for giving him the opportunity to present this research in an invited lecture.

\bibliographystyle{elsarticle-num-names} 
\bibliography{2024_refs_turb}

\end{document}